\documentclass[fleqn,10pt]{wlscirep}
\usepackage[utf8]{inputenc}
\usepackage[T1]{fontenc}
\usepackage{subcaption}
\usepackage{placeins}
\usepackage{xcolor}
\usepackage{caption}
\usepackage{tikz}

\title{AI Enhanced Cerenkov Luminescence Microscopy}

\author[1]{Stefano Pizzardi}
\author[1,2]{Lisa Alborghetti}
\author[1]{Caterina Lomazzi}
\author[1,*]{Antonello Enrico Spinelli}

\affil[1]{IRCCS San Raffaele Scientific Institute, Experimental Imaging Centre, Milan, Italy}
\affil[2]{Dep. Diagnostics and Public Health, University of Verona, Verona, Italy}
\affil[*]{corresponding author: spinelli.antonello@hsr.it}

\begin{abstract}
Cerenkov Luminescence Microscopy (CLM) enables the visualization of radiotracer uptake in three-dimensional (3D) tumor spheroids but is limited by low photon statistics and spatial blurring caused by positron transport. Here, we present a Monte Carlo (MC)-guided deep learning framework for CLM image restoration. Using TOPAS/Geant4, we developed a physical model of heterogeneous 3D tumor spheroids to simulate positron transport, Cerenkov photon generation and optical propagation. A dataset of 15,360 paired degraded and idealized ground-truth (GT) images was generated from 80 simulated spheroid geometries and used to train a supervised Content-Aware Image Restoration (CARE) network. Compared with the self-supervised Noise2Void (N2V) denoising method, CARE achieved substantially lower reconstruction errors on synthetic validation data, reducing the mean squared error from $0.0492$ to $0.0051$. Training stability and noise robustness analyses further demonstrated high reproducibility of the restored images. The trained CARE model was subsequently applied to experimental CLM images acquired from 27 U87 tumor spheroids following [$^{18}$F]FDG uptake. The restored images exhibited more localized intensity distributions and higher local contrast and signal-to-noise ratios compared with raw and N2V-processed images. These findings, aligned with the 3Rs principles of animal experimentation, demonstrate the potential of physics-based supervised restoration to mitigate both noise and positron-induced spatial broadening in CLM.
\end{abstract}

\begin{document}

\flushbottom
\maketitle
\thispagestyle{empty}

\section{Introduction}

\begin{figure}[t]
    \centering
    \includegraphics[width=\textwidth]{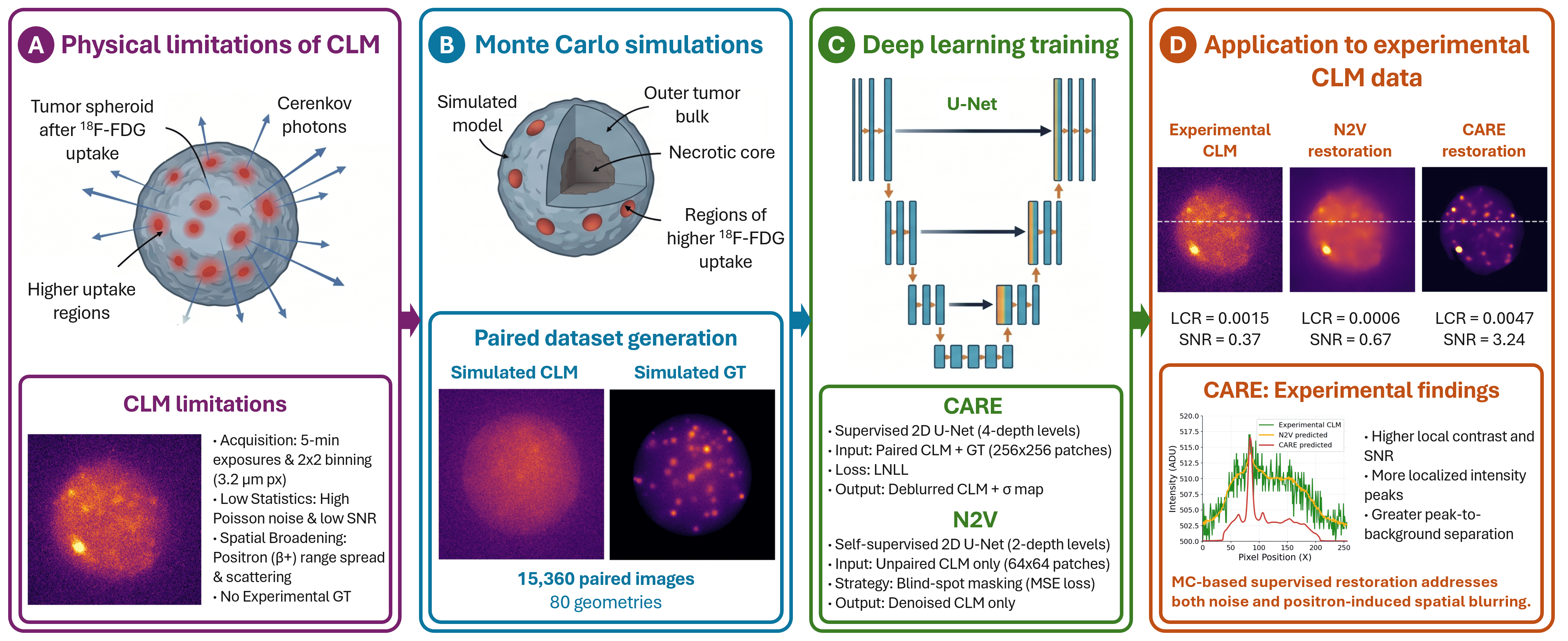}
    \caption{Overview of the MC-based deep-learning framework for CLM restoration.
(a) Physical limitations of CLM, including low photon statistics, image noise and spatial blurring caused by positron transport. Experimental GT images free from positron-induced blurring are not available, preventing direct supervised training on experimental data.
(b) MC simulations are used to model positron transport and Cerenkov photon generation in heterogeneous tumor spheroids. Paired datasets are generated by simulating CLM images and corresponding idealized GT references, in which optical photons originate directly from radionuclide decay locations, eliminating positron-induced spatial spreading.
(c) The resulting paired synthetic dataset is used to train CARE, a supervised probabilistic U-Net designed to reduce image noise and positron-induced spatial blurring. N2V, a self-supervised denoising method trained without GT references, is included for comparison.
(d) Application of the trained networks to experimental CLM acquisitions, illustrating the different effects of N2V and CARE on image intensity distributions. CARE produces higher local contrast and SNR, more localized intensity peaks and greater peak-to-background separation at the analyzed locations. Reported LCR and local SNR values represent averages across the experimental hotspot analysis.}
    \label{fig:paper_overview}
\end{figure}

Ultra-low-light microscopy is used to study biological processes, tumor metabolism and radionuclide distribution at the cellular level. Rather than relying on external excitation like conventional fluorescence, these techniques detect the intrinsically weak optical signals typically produced by bioluminescence or radionuclide decay. In order to deal with such low photon emission yield, a light-tight environment and a highly sensitive, low-noise detector \cite{ sanchez_photon-sparse_2021} are required. Common techniques include Bioluminescence Microscopy (BLM) \cite{yao_multiplexed_2022, morales-curiel_volumetric_2022} and Radioluminescence Microscopy (RLM) \cite{wang_silico_2018, khan_high-resolution_2021,pratx_radioluminescence_2012,khan_high-resolution_2021-1}. Recently, Cerenkov Luminescence Microscopy (CLM) \cite{alborghetti_cerenkov_2025} has been developed as a possible complementary approach to RLM for the imaging of different radionuclides. This technique is based on the detection of Cerenkov radiation emitted when charged particles, such as positrons ($\beta^+$) generated during radionuclide decay, travel through a dielectric medium at a velocity exceeding the local phase velocity of light. Above this Cerenkov kinetic energy threshold, the transient polarization induced by the passage of the charged particle results in electromagnetic fields that interfere constructively, producing optical photons \cite{cherenkov_visible_1934, cerenkov_visible_1937, jelley_cerenkov_1955}. Because the resulting optical signal is proportional to the radiotracer distribution without requiring any external excitation source, Cerenkov light emission has long been exploited in Cerenkov Luminescence Imaging (CLI) for macroscopic preclinical applications \cite{spinelli_cerenkov_2009,boschi_vivo_2011} and human imaging \cite{spinelli_first_2013}. Recently, Alborghetti et al.\ \cite{alborghetti_cerenkov_2025} extended the use of CLI to the microscopic scale, demonstrating the feasibility of detecting Cerenkov emission to image [$^{18}$F]FDG uptake in three-dimensional (3D) glioblastoma (GBM) spheroids. The use of CLM opens new possibilities for studying radionuclide uptake heterogeneity, tumor metabolism and radiopharmaceutical distribution with an effective pixel size of a few microns. From a translational perspective, CLM could be used as an imaging tool for \textit{in vitro} pharmacological and radiotherapy (RT) screening. More precisely, CLM performed in combination with a diagnostic radiopharmaceutical can then visualize, at microscopic resolution, any possible treatment effects. For example, it has been shown that CLM was able to detect differences of [$^{18}$F]FDG uptake in GBM spheroids treated with RT \cite{alborghetti_cerenkov_2025}. One of the advantages of CLM lies in the use of 3D \textit{in vitro} models rather than live animals. This approach is thus in line with current ethical guidelines promoting the Replacement, Reduction and Refinement (3Rs) of animal experimentation in scientific research \cite{hubrecht_3rs_2019}.\\
Despite its potential, CLM is limited by two physical effects linked to the production of Cerenkov radiation within the sample. The light yield of Cerenkov emission is low and, thus, CLM requires exposure times on the order of minutes to collect an image \cite{alborghetti_cerenkov_2025}. The low photon emission of CLM necessitates the use of pixel binning, which might limit the image spatial resolution. Secondly, because Cerenkov photons are emitted along the travel path of the charged particle, the detected optical signal is a slightly  blurred version of the true radiotracer spatial distribution \cite{alborghetti_cerenkov_2025}. In order to account for these physical effects, the use of computational image restoration algorithms, alongside hardware optimization, is needed. Classical filtering algorithms, such as Gaussian smoothing, wavelet thresholding or non-local means \cite{witkin_scale-space_1987,donoho_-noising_1995, mallat_wavelet_2008, buades_non-local_2011,gonzalez_digital_2018}, tend to suppress high-frequency details along with noise and are fundamentally unsuited to recover physical blurring. In contrast, deep learning (DL) approaches have shown substantial capabilities for low-light microscopy \cite{volpe_roadmap_2026,laine_imaging_2021}. Modern architectures, including convolutional and residual networks \cite{ronneberger_u-net_2015,zhang_beyond_2017}, generative adversarial networks \cite{goodfellow_generative_2020}, transformer-based models and diffusion probabilistic models \cite{ho_denoising_2020,  osuna-vargas_denoising_2025}, have been increasingly applied to low-light and photon-limited imaging. More specifically, frameworks such as Content-Aware Image Restoration (CARE) \cite{weigert_content-aware_2018} and networks tailored to volumetric bioluminescence \cite{morales-curiel_volumetric_2022} have successfully recovered high-resolution biological structure from acquisitions with a low Signal-to-Noise Ratio (SNR).\\
In this work, we present a simulation-driven DL framework to overcome the fundamental limits of CLM mentioned above, as shown in Fig.\ref{fig:paper_overview}. Because the experimental ground-truth (GT) images without positron blur cannot be acquired, we utilize Monte Carlo (MC) simulations to explicitly model the physics of positron transport and Cerenkov photon generation, producing matched pairs of synthetic noisy and GT images. These paired data are then used to train a fully supervised probabilistic restoration network (CARE) capable of simultaneous denoising and spatial deconvolution. To show the impact of accurate simulated training data on CLM improvement, we evaluate this approach against Noise2Void (N2V) \cite{krull_noise2void_2019,broaddus_removing_2020}, which is a self-supervised, denoising-only network.

\section{Materials and Methods}

\subsection{Experimental setup and CLM imaging}

\begin{figure}[t]
    \centering
    \begin{tikzpicture}[font=\sffamily, thick]
        
        \definecolor{daybg}{RGB}{240, 245, 250}
        \definecolor{dayborder}{RGB}{40, 90, 160}
        \definecolor{rtcolor}{RGB}{220, 50, 50}

        \draw[->, >=stealth, line width=2pt, color=gray!70] (0,0) -- (15.5,0) node[right, text=black] {\textbf{Timeline}};

        \filldraw[fill=daybg, draw=dayborder, rounded corners, line width=1pt] (0.5, 1.2) rectangle (3.2, 2.8);
        \node[align=center] at (1.85, 2.0) {\textbf{Day 0}\\U87mg Seeding\\(ULA plates)};
        \draw[->, line width=1.5pt, color=dayborder] (1.85, 1.2) -- (1.85, 0.2);
        \filldraw[fill=dayborder, draw=none] (1.85,0) circle (3.5pt);
        \node[below, font=\bfseries] at (1.85,-0.2) {Day 0};

        \filldraw[fill=daybg, draw=dayborder, rounded corners, line width=1pt] (3.8, 1.2) rectangle (6.5, 2.8);
        \node[align=center] at (5.15, 2.0) {\textbf{Day 2}\\Radiotherapy\\(20 Gy X-rays)};
        \draw[->, >=stealth, line width=1.5pt, color=rtcolor, dashed] (5.15, 3.8) -- (5.15, 2.8) node[midway, right] {\small RT Cohort};
        \draw[->, line width=1.5pt, color=dayborder] (5.15, 1.2) -- (5.15, 0.2);
        \filldraw[fill=dayborder, draw=none] (5.15,0) circle (3.5pt);
        \node[below, font=\bfseries] at (5.15,-0.2) {Day 2};

        \filldraw[fill=daybg, draw=dayborder, rounded corners, line width=1pt] (7.1, 1.2) rectangle (10.2, 2.8);
        \node[align=center] at (8.65, 2.0) {\textbf{Day 6}\\Glucose\\Starvation};
        \draw[->, line width=1.5pt, color=dayborder] (8.65, 1.2) -- (8.65, 0.2);
        \filldraw[fill=dayborder, draw=none] (8.65,0) circle (3.5pt);
        \node[below, font=\bfseries] at (8.65,-0.2) {Day 6};

        \filldraw[fill=daybg, draw=dayborder, rounded corners, line width=1pt] (10.8, 1.2) rectangle (14.5, 2.8);
        \node[align=center] at (12.65, 2.0) {\textbf{Day 7}\\1. $[^{18}\text{F}]$FDG Incubation\\2. CLM Imaging};
        \draw[->, line width=1.5pt, color=dayborder] (12.65, 1.2) -- (12.65, 0.2);
        \filldraw[fill=dayborder, draw=none] (12.65,0) circle (3.5pt);
        \node[below, font=\bfseries] at (12.65,-0.2) {Day 7};

    \end{tikzpicture}
    \caption{Schematic overview of the experimental timeline. U87mg cells were seeded on Day 0 to form 3D spheroids. On Day 2, the RT cohort was irradiated with a single 20 Gy X-ray dose. On Day 6, the growth medium was replaced with a glucose-free formulation to induce metabolic starvation. On Day 7, spheroids were incubated with [$^{18}$F]FDG and subsequently imaged using CLM.}
    \label{fig:experimental_timeline}
\end{figure}
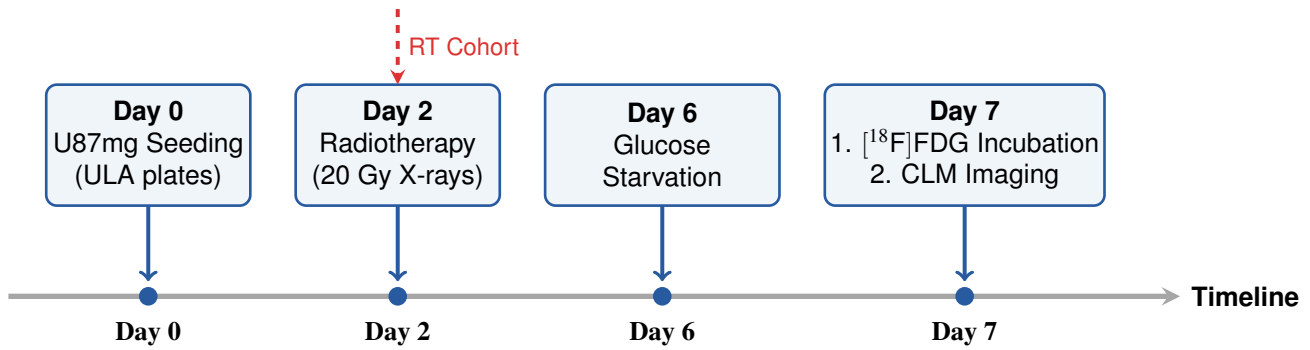

The experimental CLM spheroid dataset used in this study comprises the 10-spheroid cohort originally described by Alborghetti et al. \cite{alborghetti_cerenkov_2025} and an additional 9 irradiated and 8 control spheroids. These additional spheroids were acquired following the same experimental protocol, yielding a total of 27 spheroids (14 RT, 13 CTRL). The experimental procedures followed the methodology previously described by our group \cite{alborghetti_cerenkov_2025}. A schematic representation of the complete experimental timeline is provided in Fig.~\ref{fig:experimental_timeline}. Since the biological implications of RT on these models have already been investigated by Alborghetti et al. \cite{alborghetti_cerenkov_2025}, the present study focuses strictly on the computational enhancement of the CLM images. Consequently, no distinction was made between the treated and control groups and all 27 spheroids were pooled into a single dataset for image restoration.

\subsubsection{Cell culture and irradiation}
Experiments were conducted using U87mg Luc+ cells (a human glioma cell line) maintained in Dulbecco's Modified Eagle Medium (DMEM) supplemented with 10\% (v/v) Fetal Bovine Serum (FBS), GlutaMAX and 1\% (v/v) Penicillin-Streptomycin. To generate 3D tumor spheroids, cells were seeded at a density of $1.25 \times 10^4$ cells/mL in ultra-low attachment (ULA) 96-well round-bottom plates (200 $\mu$L/well) and kept in a humidified incubator at 37$^\circ$C with 5\% CO$_2$. On Day 2, a subset of spheroids underwent RT with a single 20 Gy X-ray dose delivered via a Small Animal Radiation Research Platform (SmART, PXI), while the remaining spheroids served as untreated controls.

\subsubsection{Radiotracer incubation and sample preparation}
On Day 6 (the day prior to imaging), the growth medium was replaced with glucose-free medium to induce metabolic starvation. On the day of CLM acquisition (Day 7), spheroids were incubated for 60 minutes at 37$^\circ$C in 200 $\mu$L of glucose-free medium supplemented with 5 MBq of [$^{18}$F]FDG. Following incubation, spheroids were washed three times (2 minutes each) in non-radioactive medium, carefully retrieved using a pipette and placed on a microscope slide within a small droplet of growth medium to prevent drying during the CLM optical acquisition. CLM imaging was then performed immediately after sample preparation.

\subsubsection{Imaging system and acquisition parameters}

The imaging system consisted of a custom light-tight box equipped with a back-illuminated EMCCD sensor (Andor iXon Ultra 897, $512 \times 512$ pixels), a 60 mm C-mount extension tube and a $20\times/0.45$ NA microscope objective. After focusing the sample using bright-field LED illumination, the light-tight box was sealed, covered with blackout fabric (Thorlabs) and room lights were turned off to ensure a dark imaging environment. To minimize thermal noise, the CCD sensor was cooled to $-80^\circ$C and operated in conventional mode (EM gain disabled) to prevent read noise amplification, also considering that a very low frame rate was required. For each of the 27 spheroids, one or two consecutive 5-minute frames were acquired, yielding a total dataset of 46 individual CLM images. The consecutive frames were not summed. Instead, each 5-minute acquisition was kept as a separate image and analyzed individually. A $2\times2$ hardware pixel binning was applied to enhance signal intensity, producing final $256 \times 256$ pixel images with an effective pixel size of 3.2 $\mu$m and a field of view of approximately $820 \times 820$ $\mu$m.

\subsection{Monte Carlo simulations}

To overcome the intrinsic lack of an experimental GT, we developed a novel MC simulation model to generate paired datasets of simulated optical images. Simulations were carried out using TOPAS version 3.9 \cite{perl_topas_2012,faddegon_topas_2020}, based on Geant4 version 10.07-patch-03. The simulations use the \texttt{g4em\_standard\_opt4} physics list for electromagnetic interactions and particle tracking, coupled with the \texttt{g4optical} physics list to enable Cerenkov photon generation and optical photon propagation. A standard global secondary particle production cutoff (range cut) of 0.05 mm was applied, corresponding to an electron energy of approximately 50 keV in soft tissue. Since this value is safely below the Cerenkov emission threshold ($E_{thr} \approx 250$ keV for $n=1.35$), this cutoff effectively reduces computational costs without compromising the simulation of Cerenkov photons. Computations were executed on a dedicated research cluster comprising 17 CPU nodes (17 × XL190r Gen10 blades), each equipped with 2 × Intel(R) Xeon(R) Gold 6240 CPUs and 384 GB of RAM.

\subsubsection{Spheroid geometry}

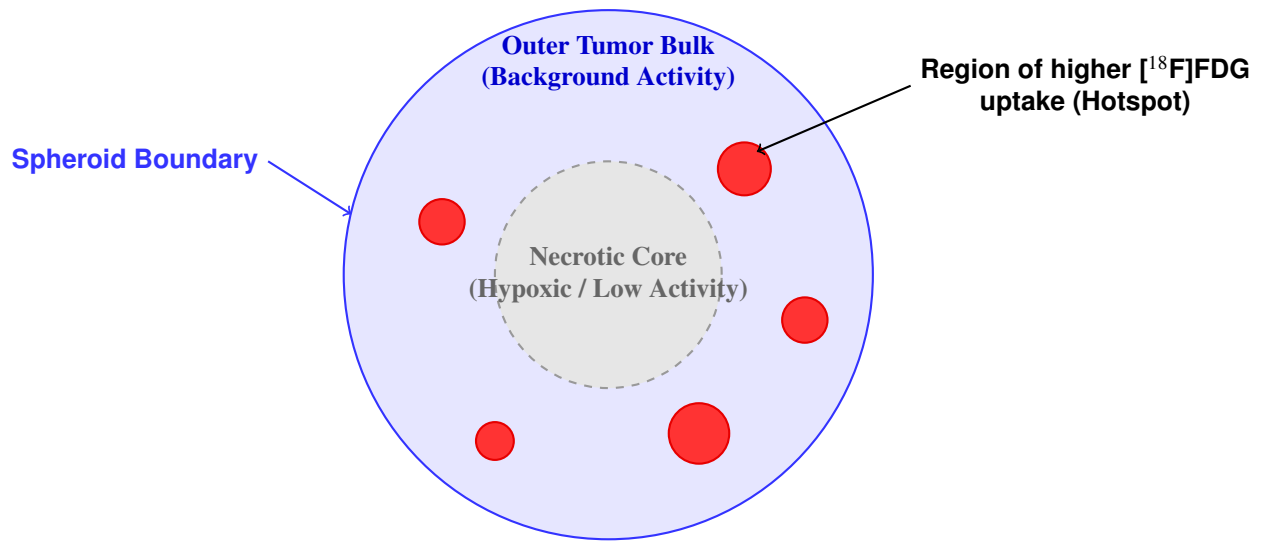
\begin{figure}[t]
    \centering
    \begin{tikzpicture}[font=\sffamily]
        
        \draw[fill=blue!10, draw=blue!80, thick] (0,0) circle (3.5cm);
        \node[text=blue!80!black, font=\bfseries, align=center] at (0, 2.8) {Outer Tumor Bulk\\(Background Activity)};
        
        \draw[fill=gray!20, draw=gray!80, thick, dashed] (0,0) circle (1.5cm);
        \node[text=gray!80!black, font=\bfseries, align=center] at (0, 0) {Necrotic Core\\(Hypoxic / Low Activity)};
        
        \fill[red!80, draw=red!90!black, thick] (1.8, 1.4) circle (0.35cm);
        \fill[red!80, draw=red!90!black, thick] (-2.2, 0.7) circle (0.3cm);
        \fill[red!80, draw=red!90!black, thick] (1.2, -2.1) circle (0.4cm);
        \fill[red!80, draw=red!90!black, thick] (-1.5, -2.2) circle (0.25cm);
        \fill[red!80, draw=red!90!black, thick] (2.6, -0.6) circle (0.3cm);
        
        \draw[<-, thick, shorten <= 2pt] (1.9, 1.6) -- (4.0, 2.5) 
            node[right, align=center] {\textbf{Region of higher [$^{18}$F]FDG}\\ \textbf{uptake (Hotspot)}};
            
            
        \draw[<-, thick, blue!80] (-3.4, 0.8) -- (-4.5, 1.5) 
            node[left] {\textbf{Spheroid Boundary}};

    \end{tikzpicture}
    \caption{Schematic 2D cross-section of the simulated 3D tumor spheroid geometry. The model comprises an outer viable tumor bulk, an inner hypoxic necrotic core and randomly distributed hotspots representing local clusters of higher [$^{18}$F]FDG uptake regions.}
    \label{fig:spheroid_geometry_schema}
\end{figure}

The 3D tumor spheroid was modeled as a spherical volume with a radius ranging from 200 to 350~$\mu$m. Its mass density was set to 1 g/cm$^3$ and its elemental composition followed the standard ICRU four-element tissue definition \cite{mcnair_icru_1981}: H (10.1 wt\%), C (11.1 wt\%), N (2.6 wt\%) and O (76.2 wt\%). As shown in Fig. \ref{fig:spheroid_geometry_schema}, the spheroid volume was divided into three regions with distinct simulated activity: an inner necrotic core, an outer tumor bulk and a set of higher [$^{18}$F]FDG uptake regions (hereafter referred to as hotspots for simplicity). The necrotic core represented the central hypoxic region where cell viability and metabolic activity are severely diminished. The surrounding tumor bulk (the outer, still-viable shell of the spheroid) represented background activity. To simulate regions with elevated [$^{18}$F]FDG uptake, multiple hotspots modeled as non-overlapping spherical sub-volumes were distributed throughout the bulk volume.\\
To maximize dataset variance and prevent the neural network from overfitting to fixed structural patterns, a custom Python script was developed to dynamically generate unique spheroid geometries for each simulation run. Rather than relying on fixed dimensions, structural parameters (e.g. the radius of the spheroid, necrotic core and hotspots, along with the hotspot positions and their relative radiotracer uptake intensities) were stochastically generated. For each parameter, the script selected an initial center value and subsequently sampled the final simulation value from a log-normal distribution centered around it. A total of 80 such stochastically varied geometries were generated, forming the basis of the full training dataset.\\
To closely replicate the physical imaging conditions, the generated spheroid was virtually immersed in a surrounding spherical volume (radius 2.5 mm) representing a droplet of biological culture medium. In the experimental setup, placing the spheroid in a droplet of medium is necessary to prevent tissue dehydration during the prolonged exposure times required for CLM imaging. The surrounding culture medium (DMEM + 10\% FBS) was modeled as a homogeneous material with a mass density of 1.01 g/cm$^3$ and an effective elemental composition of H (10.8 wt\%), O (87.5 wt\%), C (0.5 wt\%), Na (0.4 wt\%), Cl (0.4 wt\%) and N (0.4 wt\%), approximating the composition of an aqueous cell culture medium \cite{poon_measuring_2022,moore_nanoparticle_2015,chou_removal_2015}.

\subsubsection{Activity modeling and paired CLM-GT simulation}
The three defined geometric regions (necrotic core, outer bulk and hotspots) were configured as distinct, independent volumetric radiation sources within the TOPAS MC code. Because explicitly simulating the decay of individual cells is computationally very demanding, emission densities were assigned on a regional volumetric basis. The relative activity, and therefore the simulated tracer uptake, was controlled by modulating the number of primary particles generated within each region. The Python script assigned a low emission density to the necrotic core, a baseline density to the outer tumor bulk and significantly elevated emission densities to the randomly generated hotspots. To reflect realistic biological heterogeneity, this elevated activity was not uniform across all hotspots. Instead, each individual hotspot was assigned a distinct, randomly varied emission density. For each simulated spheroid, the total number of primary $^{18}$F decay events across all source regions was randomly sampled between $6\times10^8$ and $1.2\times10^9$. In the MC model, each simulated primary particle represents a single $^{18}$F decay, and the subsequent positron transport and Cerenkov photon generation are simulated explicitly. The range of $6\times10^8$-$1.2\times10^9$ primary events was selected to produce simulated images spanning a broad range of photon-statistics conditions, from qualitatively noisier to less noisy than those observed experimentally. Randomizing the total number of primary events across simulations introduced additional variability in the MC counting statistics and avoided training the network on a single fixed photon-statistics regime.\\
For each specific spheroid geometry, two matched simulations were run. In the first, primary particles (positrons) were emitted uniformly and isotropically from the source volumes, with initial kinetic energies sampled from the $\beta^+$ emission spectrum of $^{18}$F. As these positrons traveled through the spheroid and culture medium, they generated Cerenkov photons along their tracks whenever their kinetic energy exceeded the local production threshold, so that the resulting noisy image captures both the physical spatial blurring intrinsic to positron transport (the "positron spread") and the low photon statistics characteristic of experimental CLM acquisitions. In the second simulation, $2.4\times10^9$ optical photons were generated isotropically as primary particles directly within the same source volumes, preserving the relative activity distribution across regions. The large number of primary photons minimized statistical fluctuations, while their direct emission from the source volumes eliminated positron-induced spatial broadening. Since these photons still underwent optical scattering and absorption, the resulting GT image provided a low-noise and idealized optical reference of the simulated activity distribution.

\subsubsection{Optical properties and boundary conditions}

\begin{table}[t]
\centering
\caption{Optical parameters used in the TOPAS MC simulations.}
\label{tab:optical_parameters}

\vspace{0.2cm}
\textbf{(a) 3D Tumor Spheroid Model} \\
\vspace{0.1cm}
\begin{tabular}{ccccc}
\toprule
\textbf{Wavelength (nm)} & \textbf{Energy (eV)} & \textbf{Mean Scattering Path (mm)} & \textbf{Mean Absorption Path (mm)} & \textbf{Anisotropy Factor ($g$)} \\
\midrule
350 & 3.54 & 0.025 & 1.650 & 0.875 \\
400 & 3.10 & 0.035 & 2.185 & 0.875 \\
450 & 2.76 & 0.043 & 4.383 & 0.875 \\
500 & 2.48 & 0.049 & 7.465 & 0.875 \\
600 & 2.07 & 0.065 & 11.850 & 0.875 \\
700 & 1.77 & 0.083 & 22.154 & 0.875 \\
\bottomrule
\end{tabular}

\vspace{0.6cm} 

\textbf{(b) Culture Medium (DMEM + 10\% FBS)} \\
\vspace{0.1cm}
\begin{tabular}{ccccc}
\toprule
\textbf{Wavelength (nm)} & \textbf{Energy (eV)} & \textbf{Mean Scattering Path (mm)} & \textbf{Mean Absorption Path (mm)} & \textbf{Anisotropy Factor ($g$)} \\
\midrule
350 & 3.54 & 150 & 10.49 & 0 \\
400 & 3.10 & 150 & 11.12 & 0 \\
450 & 2.76 & 150 & 9.60  & 0 \\
500 & 2.48 & 150 & 11.80 & 0 \\
600 & 2.07 & 150 & 46.40 & 0 \\
700 & 1.77 & 150 & 194.92 & 0 \\
\bottomrule
\end{tabular}
\end{table}

To simulate photon transport through the biological sample, wavelength-dependent optical properties were assigned to both the tumor spheroid and the surrounding culture medium. Scattering in the spheroid was modeled using a Mie-based description of photon interactions with biological structures whose characteristic dimensions are comparable to or larger than the wavelength of visible light \cite{jacques_optical_2013}.\\
Since exact optical properties for U87 multicellular tumor spheroids are not available across the full spectral range considered here, reference values were derived from experimental measurements reported for biologically related tumor tissues, including human astrocytoma, glioblastoma and early-stage glioma models \cite{yaroslavsky_optical_2002, genina_optical_2019, honda_determination_2018, jacques_optical_2013, hargrave_optical_1996}. To minimize inconsistencies arising from the different optical-transport models and parameterizations used in individual studies, the reduced scattering coefficient, $\mu_s'$, was adopted as the common transport metric. At each wavelength, the available literature values of $\mu_s'$ were averaged to obtain a representative wavelength-dependent reduced scattering coefficient for the spheroid. The corresponding scattering mean free path was then calculated as $l_s = (1-g)/\mu_s'$, where $g$ is the anisotropy factor. The refractive index of the spheroid was set to $1.365$, consistent with optical coherence tomography measurements of multicellular tumor spheroids \cite{hari_optical_2019}. Since the spheroids are avascular, absorption coefficients exhibiting pronounced hemoglobin-related spectral features were considered less representative of the in vitro model. Accordingly, absorption values were selected to minimize the contribution of hemoglobin-related absorption.\\
The culture medium (DMEM + 10\% FBS) was modeled as a weakly scattering aqueous medium with a refractive index of $1.35$ \cite{hoang_optical_2019, barroso_refractive_2019}. The wavelength-dependent absorption lengths were derived from UV-Vis absorbance measurements of the supplemented medium \cite{ferencik_pa12_2024}, treating the measured attenuation as an upper-bound estimate of the absorption contribution. Scattering was approximated as isotropic ($g=0$), with a constant effective scattering mean free path of $l_s=150$~mm. Given the short photon propagation distances relative to this mean free path, scattering within the medium was expected to have a limited contribution to spatial broadening compared with the spheroid. The optical parameters are summarized in Table~\ref{tab:optical_parameters}.\\
Finally, to avoid idealized spherical symmetry and numerical artifacts associated with perfectly specular interfaces in Geant4, a small surface roughness was introduced at the spheroid-medium boundary using the UNIFIED optical model, with $\texttt{SigmaAlpha}=0.2$~rad. 

\subsubsection{Optical scoring, data augmentation and detector noise modeling}
To efficiently generate a robust and highly variable dataset from these computationally intensive simulations, a multi-view phase-space scoring geometry was implemented. A number of 24 distinct phase-space scoring planes, each measuring 500 × 500 $\mu$m$^2$, were arranged cylindrically at a radial distance of 4 mm from the spheroid center, positioned at 15-degree intervals. The optical photons crossing these planes were recorded alongside their physical coordinates, direction vectors and energies.\\
Because the internal simulated radiotracer distribution, due to the randomized hotspots, was inherently asymmetric, each of the 24 viewing angles yielded a geometrically unique 2D projection. This multi-view strategy effectively functioned as a physics-based data augmentation technique, multiplying the dataset yield per simulation run by a factor of 24.\\
Photon transport from the phase-space scoring planes to the virtual sensor was handled in post-processing via a custom Python script. Transport from the scoring plane to a virtual EMCCD sensor was simulated using a ray-transfer-matrix approach based on 6 × 6 optical matrices \cite{lin_66_2009,lin_matrix-based_2006}. In this formalism, the optical system is represented as a sequence of matrix operations acting on the photon coordinates ($x, y, z, dx, dy, dz$), where $x, y, z$ denote spatial coordinates and $dx, dy, dz$ the components of the direction vector. An ideal thin lens with a focal length of $f = 2$ mm was simulated to focus the central plane of the phantom onto a virtual sensor placed 4 mm from the phase-space plane. Together with the 4 mm phase-space-to-spheroid distance defined above, this corresponds to both the object and image distances being set to $2f$, yielding a 1:1 magnification. The resulting 2D optical image provides a simplified representation of the photon distribution incident on the experimental imaging system. It should be noted that this ideal thin lens approximation does not account for real imaging distortions, which were not explicitly modeled in the simulation.\\
Following photon transport to the virtual sensor plane, each image was further augmented via eight rigid geometric transformations (four 90 degree rotations, each with and without a horizontal flip), applied identically to the paired noisy and GT images so as to strictly preserve pixel-wise correspondence. Combined with the 24-view phase-space scoring geometry, this generated 192 paired training images per simulated spheroid geometry ($24 \times 8$), for a total of 15,360 paired images across the 80 simulated geometries.\\
Because the MC simulations provide optical photon counts at the virtual sensor plane, whereas the experimental CLM images are recorded as camera output values in analog-to-digital units (ADU), a direct quantitative conversion between the simulated signal and the experimental intensity scale could not be established without a dedicated calibration of the camera response. Rather than imposing a fixed conversion factor, we adopted an intensity-scaling procedure. For each simulated noisy image, the MC signal was divided by a randomly sampled factor between 4 and 10. The range was selected so that the resulting images covered a broad range of signal intensities, including conditions qualitatively weaker and stronger than those observed in the 5-minute experimental CLM acquisitions. This random scaling was combined with the variability introduced by the number of simulated $^{18}$F decay events, thereby exposing the network to a broad distribution of signal and photon-statistics conditions rather than a single fixed intensity level.\\
To reproduce the experimental camera background, multiple 5-minute acquisitions were also collected with the shutter closed under the same imaging conditions. These dark frames exhibited a reproducible background pedestal in the experimental ADU images. The mean and standard deviation of the pixel values across the collected dark frames were used to characterize the experimental background distribution and a stochastic Gaussian background contribution was subsequently added to each rescaled MC image. This procedure was intended to reproduce the first-order statistical characteristics of the experimental background and detector noise.\\
The dataset was split into training and validation sets at the level of individual spheroid geometries rather than individual images: all 192 augmented images belonging to a given geometry were assigned entirely to either the training or the validation partition. This spheroid-level split, 70 geometries (13,440 images, 87.5\%) for training and 10 geometries (1,920 images, 12.5\%) for validation, was adopted specifically to prevent data leakage, since near-duplicate augmented views of the same underlying geometry would, if split across partitions, artificially inflate validation performance and invalidate model selection.

\subsection{Neural networks and AI training}

We implemented a deep learning image restoration pipeline based on U-Net convolutional neural networks \cite{ronneberger_u-net_2015}. CARE was used as a fully supervised restoration network, trained exclusively on the paired synthetic MC dataset using simulated noisy Cerenkov images as input and the corresponding direct-emission optical images as GT targets. The trained CARE model was subsequently applied to both synthetic and experimental data. N2V was used as a self-supervised denoising baseline that does not require paired GT images \cite{krull_noise2void_2019,broaddus_removing_2020}. Two N2V models were trained independently: one on the synthetic MC dataset and one directly on the experimental CLM acquisitions. This design allowed comparison of the physics-informed CARE restoration with a denoising-only approach under both synthetic and experimental training conditions. All network training was performed using an NVIDIA GB10 GPU with 128 GB of memory.

\subsubsection{Supervised restoration: CARE}

The network was configured as a 2D U-Net with four downsampling levels, $3\times3$ convolutional kernels and 64 feature channels in the first level, doubling at each subsequent level. The resulting model contained 21,976,578 trainable parameters. The default nearest-neighbor upsampling layers were replaced with bilinear interpolation to avoid blocking artifacts.\\
A probabilistic Laplace negative log-likelihood (LNLL) loss was used, with two output channels per pixel corresponding to the predicted reconstruction and scale parameter $\sigma$. This probabilistic formulation provided spatially resolved estimates of reconstruction uncertainty in addition to the restored image. During the initial stages of training, the LNLL loss was spatially weighted to prioritize spheroid pixels (defined by a GT intensity threshold of 0.3) over the background. This spatial weighting was progressively attenuated across training steps (CSBDeep decay factor of 0.01), allowing the training criterion to gradually transition toward a spatially uniform loss.\\
The 15,360 paired MC images were divided into 13,440 training images (70 geometries) and 1,920 validation images (10 geometries). Each image was used as a full $256\times256$ training patch. The model was optimized using Adam with a batch size of 32, 400 steps per epoch and an initial learning rate of $4\times10^{-4}$. The learning rate was reduced by a factor of 0.5 after 20 epochs without improvement in validation loss. Training was limited to 400 epochs and early stopping was applied with a patience of 40 epochs, restoring the weights corresponding to the best validation loss.\\
To improve robustness to residual differences between the simulated and experimental domains, additional on-the-fly augmentation was applied during training. Input intensities were randomly scaled between 85\% and 115\%, followed by additive Gaussian noise with a standard deviation randomly sampled between 2\% and 5\% of the maximum intensity of the scaled image.

\subsubsection{Self-supervised denoising: N2V}

N2V was used as a self-supervised denoising baseline based on a blind-spot training strategy, in which selected pixels are masked and predicted from their surrounding spatial context \cite{krull_noise2void_2019,broaddus_removing_2020}. Under the assumption that the noise is not spatially correlated with neighboring pixels, this allows the network to learn image structure without requiring clean GT targets. Unlike CARE, N2V was not provided with an explicit representation of the unblurred tracer distribution and was therefore used exclusively as a denoising approach.\\
For both N2V models, a 2D U-Net architecture with $3\times3$ convolutional kernels, 96 feature channels in the first level, two U-Net depth levels and no residual formulation was used. Batch normalization was enabled and the mean squared error (MSE) loss was optimized using Adam. Blind-spot masking was configured with $n2v\_perc\_pix=0.198$, corresponding to approximately 8 manipulated pixels per $64\times64$ patch. Pixels were manipulated using the \texttt{uniform\_withCP} strategy with a neighborhood radius of 2 pixels. The same architecture and blind-spot parameters were used for both training domains.\\
The synthetic N2V model was trained using the same 15,360 simulated CLM images used for CARE. The dataset was split at the spheroid-geometry level, with 70 geometries assigned to training and 10 to validation, corresponding to 13,440 and 1,920 images, respectively. Patch extraction produced 107,520 training patches and 15,360 validation patches of $64\times64$ pixels. Training used a batch size of 128, 840 steps per epoch and a maximum of 200 epochs, with an initial learning rate of $4\times10^{-4}$. The learning rate was reduced by a factor of 0.5 after 10 consecutive epochs without improvement in validation loss. Early stopping with a patience of 20 epochs was also applied, and the weights corresponding to the lowest validation loss were restored.\\
The second N2V model was trained directly on the experimental CLM acquisitions in order to provide a domain-specific denoising baseline. The dataset comprised all 46 available raw experimental CLM frames, corresponding to one or two 5-minute acquisitions per spheroid. All available frames were retained as separate training examples rather than averaged or combined. In particular, consecutive frames from the same spheroid were treated as distinct acquisitions because they represented distinct realizations of the measured signal and detector noise, with the second acquisition also potentially exhibiting reduced signal due to the ongoing radioactive decay. A total of 5,888 $64\times64$ patches were extracted and randomly divided into 5,300 training patches and 588 validation patches (90/10 split). Because the split was performed at the patch level, patches from the same frame could occur in both partitions; consequently, the validation subset was used for monitoring training progress rather than as an independent image-level test set. Training used a batch size of 128, 100 steps per epoch and a maximum of 200 epochs, with an initial learning rate of $4\times10^{-4}$ reduced by a factor of 0.5 after 10 consecutive epochs without improvement in validation loss. No early-stopping callback was used, and training therefore continued for the full 200 epochs. The checkpoint with the lowest validation loss was subsequently used for prediction.\\

\subsection{Evaluation metrics and comparative analysis}

To assess structural similarity and reproducibility, the Structural Similarity Index (SSIM) \cite{wang_image_2004} and Fourier Ring Correlation (FRC) \cite{nieuwenhuizen_measuring_2013,van_heel_fourier_2005} were employed. SSIM quantifies image similarity by comparing local intensity, contrast and structural information between two images and is defined as
\begin{equation}
\mathrm{SSIM}(x,y) = \frac{(2\mu_x\mu_y+C_1)(2\sigma_{xy}+C_2)}{(\mu_x^2+\mu_y^2+C_1)(\sigma_x^2+\sigma_y^2+C_2)},
\end{equation}
where $\mu_x$ and $\mu_y$ are the local mean intensities, $\sigma_x^2$ and $\sigma_y^2$ are the local variances, $\sigma_{xy}$ is the local covariance, and $C_1$ and $C_2$ are numerical stabilization constants. FRC measures the correlation between two images in the Fourier domain as a function of spatial frequency $q$, and is defined as
\begin{equation}
\mathrm{FRC}(q) = \frac{\displaystyle\sum_{\mathbf{k}\in R_q} F_1(\mathbf{k})F_2^*(\mathbf{k})}{\sqrt{\displaystyle\sum_{\mathbf{k}\in R_q}|F_1(\mathbf{k})|^2\displaystyle\sum_{\mathbf{k}\in R_q}|F_2(\mathbf{k})|^2}},
\end{equation}
where $F_1$ and $F_2$ are the Fourier transforms of the two images, $^*$ denotes complex conjugation and $R_q$ represents the set of Fourier coefficients within a ring of spatial frequency $q$.\\
Since experimental CLM does not provide repeated acquisitions of the same physical sample, these analyses were performed on synthetic data only. SSIM was computed using the \texttt{scikit-image} implementation with $7\times7$ sliding windows after min-max normalization. FRC was computed using the dedicated Fiji/ImageJ plugin and the effective resolution limit was defined as the spatial frequency at which the FRC curve fell below the conventional $1/7$ threshold. For training-stability analysis, SSIM and FRC were computed between the predictions of two independently trained CARE models applied to the same validation inputs. For noise-robustness analysis, SSIM and FRC were computed between pairs of simulations generated from identical spheroid geometries with independent noise realizations, both before and after CARE restoration.\\
To evaluate restoration accuracy and compare CARE against N2V, both trained models were tested on the synthetic validation set, allowing direct comparison between network predictions and the corresponding GT references. Global restoration performance was assessed using the MSE and mean absolute error (MAE). All images were independently min-max normalized to a $[0,1]$ range prior to error computation. Local recovery of sharp intensity gradients was visually assessed by extracting 1D spatial intensity profiles. For visual comparison, the extracted profiles were independently normalized using their respective 1st and 99th intensity percentiles.\\
To evaluate restoration performance on experimental samples in the absence of an experimental GT, absolute error metrics such as MSE and MAE could not be computed. Instead, two complementary metrics were considered: the Local Contrast Ratio (LCR) and local SNR. LCR and SNR were used to quantify local hotspot visibility and contrast.\\
Hotspot detection was performed exclusively on the CARE prediction. For each frame, the intensity pedestal was estimated using the pixel median background intensity. The estimated pedestal was subtracted from the CARE image. A spheroid mask was then defined from the pedestal-corrected CARE image. Candidate hotspots were identified within this mask as local maxima exceeding the intra-spheroid mean intensity by two standard deviations.\\
For each detected hotspot, a circular peak ROI ($r\leq4$ pixels) and an annular background ROI ($6<r\leq15$ pixels) were defined, with the latter restricted to the spheroid mask. The same hotspot coordinates and ROIs were applied to the experimental CLM, N2V predicted and CARE predicted images. The LCR and local SNR were calculated as
\begin{equation}
\mathrm{LCR} =
\frac{\mu_{\mathrm{hotspot}}-\mu_{\mathrm{bg}}}
{\mu_{\mathrm{bg}}},
\qquad
\mathrm{SNR} =
\frac{\mu_{\mathrm{hotspot}}-\mu_{\mathrm{bg}}}
{\sigma_{\mathrm{bg}}},
\end{equation}
where $\mu_{\mathrm{hotspot}}$ is the mean intensity within the peak ROI and $\mu_{\mathrm{bg}}$ and $\sigma_{\mathrm{bg}}$ are the mean and standard deviation within the local-background ROI.\\
For each frame, LCR and SNR values were first averaged across all detected hotspots. The resulting per-frame values were then pooled across the 27 experimental frames in which at least one hotspot was detected and are reported as mean $\pm$ standard deviation alongside their median values.

\section{Results}

\subsection{Synthetic dataset and CARE network training}

\begin{figure}[t]
    \centering
    \includegraphics[width=\textwidth]{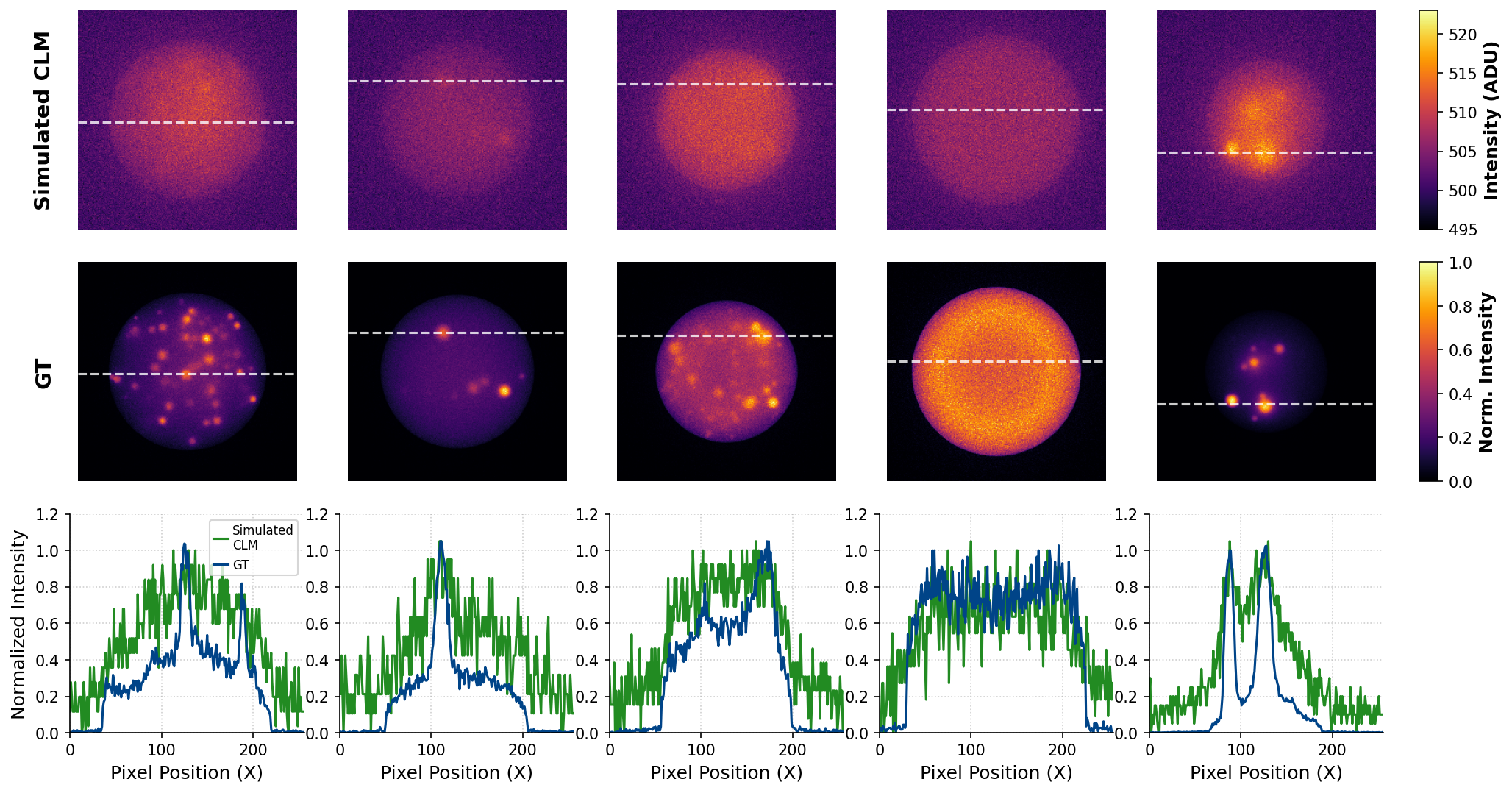}
    \caption{MC simulation dataset overview. Top row: Simulated CLM acquisitions, incorporating physical positron range spread, optical scattering, Poisson noise and detector noise. To mimic experimental conditions, these images are displayed in their raw ADU intensity scale. Middle row: Corresponding GT images representing the spatial distribution of the simulated activity, without positron-induced blur. Each GT image is independently normalized to a [0, 1] range. Dashed white lines denote the cross-sections extracted for 1D profiling. Bottom row: 1D spatial intensity profiles comparing the sharp GT signal (blue line) against the degraded simulated CLM signal (green line) across five representative 3D tumor spheroid geometries. The intensity profiles were normalized using their respective 1st and 99th percentiles.}
    \label{fig:mc_dataset}
\end{figure}

\begin{figure}[tp]
    
    \makebox[\textwidth][l]{\textbf{(a)}} \\[0mm]
    \includegraphics[width=0.9\textwidth]{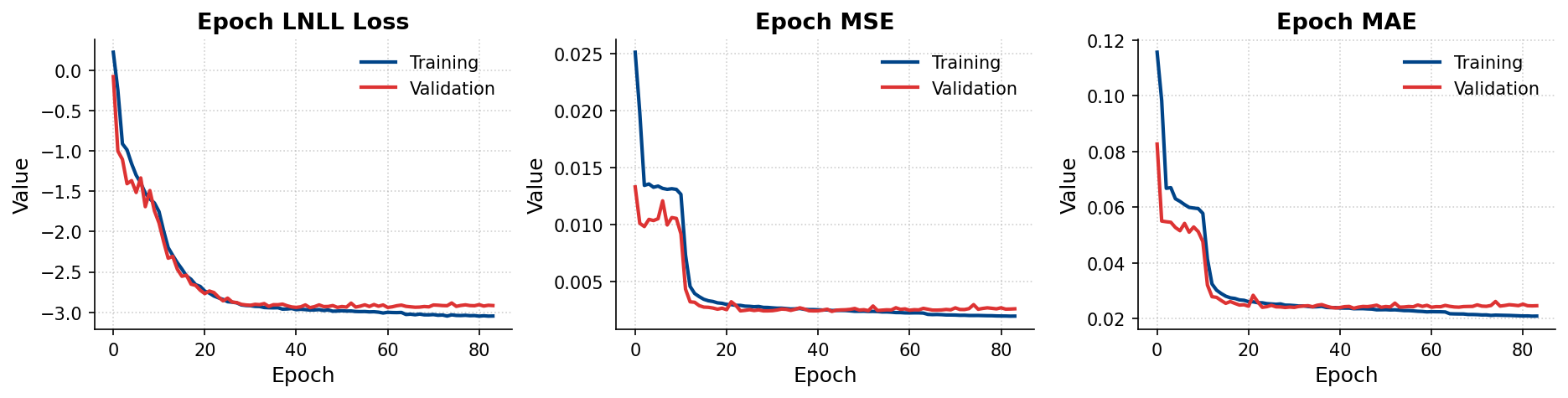} \\[-6mm]
    
    \makebox[\textwidth][l]{\textbf{(b)}} \\[0mm]
    \includegraphics[width=\textwidth]{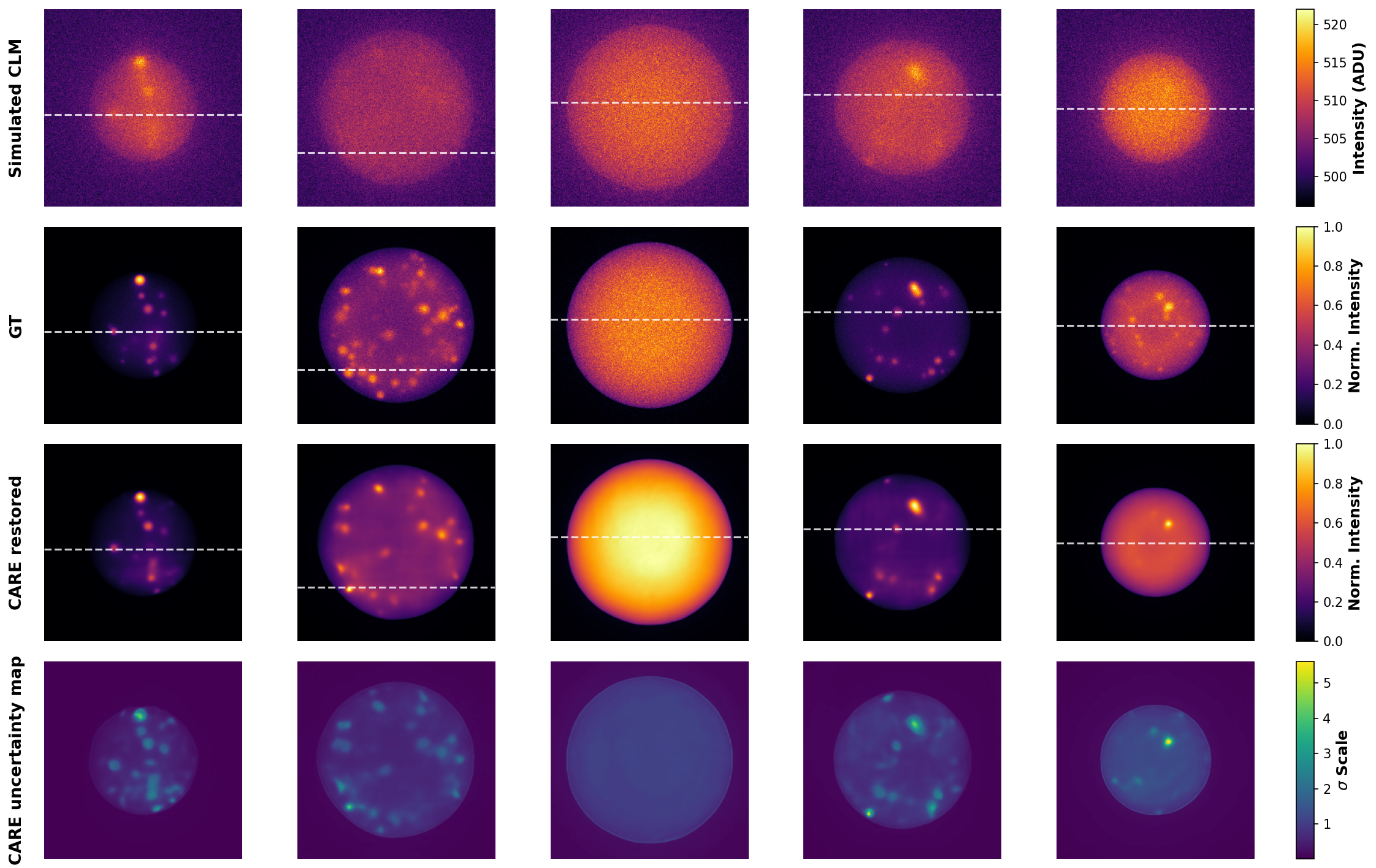} \\[-5mm]
    
    \makebox[\textwidth][l]{\textbf{(c)}} \\[0mm]
    \includegraphics[width=0.93\textwidth]{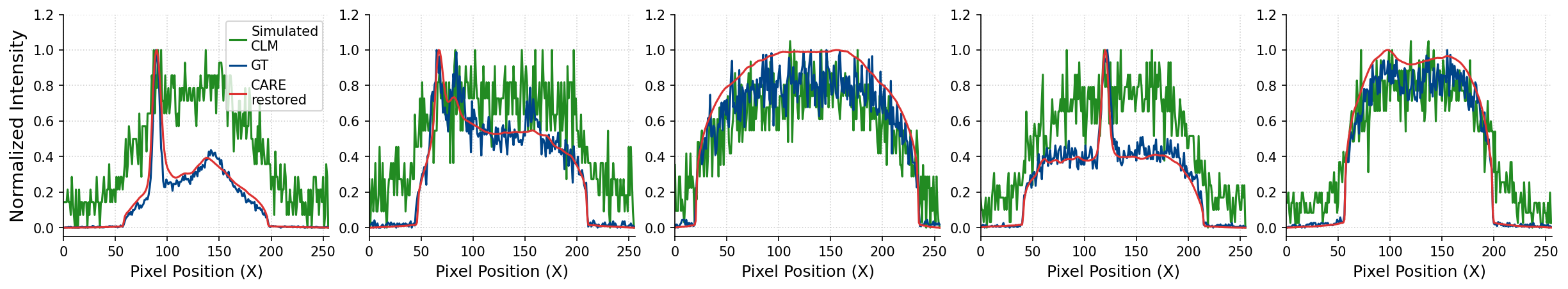}
    
    \caption{CARE network training performance. 
    (a) Training and validation metrics (LNLL loss, MSE and MAE) evaluated across epochs. 
    (b) Restoration on five representative geometries of the validation dataset. Rows display (top to bottom): simulated CLM images, GT images, CARE predictions and CARE uncertainty maps ($\sigma$). Dashed white lines indicate cross-sections used for 1D profiling. 
    (c) 1D intensity profiles comparing the CLM signal (green line), the GT signal (blue line) and the CARE restored signal (red line). To ensure a robust visual comparison, all profiles were normalized to a [0, 1] range using their respective 1st and 99th percentiles.}
    \label{fig:care_complete_results}
\end{figure}

The MC simulation pipeline generated a heterogeneous dataset of 3D tumor spheroids. Fig.~\ref{fig:mc_dataset} presents five representative examples, ranging from relatively homogeneous activity distributions to geometries containing multiple localized hotspots. Compared with the corresponding GT images, the simulated CLM images exhibit broader intensity distributions and reduced separation between neighboring hotspots. In addition, some hotspots are partially or completely obscured by noise, making them difficult to distinguish from the surrounding signal. These differences are also visible in the 1D intensity profiles (Fig.~\ref{fig:mc_dataset}, bottom row), where distinct peaks in the GT appear broadened, merged or masked by noise in the corresponding CLM profiles.\\
The paired synthetic dataset was used to train the CARE network. Training stopped at epoch 84 following the early-stopping criterion, with the optimal weights restored from epoch 44. At this epoch, the validation MSE and MAE were 0.0024 and 0.0237, respectively. The training and validation curves showed similar trends throughout the training (Fig.~\ref{fig:care_complete_results}a).\\
Fig.~\ref{fig:care_complete_results}b shows the CARE reconstructions for five representative geometries from the validation dataset. Compared with the simulated noisy inputs, the reconstructed images exhibit more localized intensity distributions and improved separation between several neighboring hotspots, approaching the spatial patterns observed in the corresponding GT images. However, some hotspots visible in the GT remain unresolved in the CARE reconstructions. The predicted uncertainty maps ($\sigma$) show low values in the image background and generally higher values within the spheroid, particularly at the hotspot locations.\\
The corresponding 1D intensity profiles (Fig.~\ref{fig:care_complete_results}c) further illustrate the differences between the simulated CLM images, CARE predictions and GT. CARE predictions profiles generally reproduce the positions and separation of prominent GT peaks more closely than the simulated CLM profiles, although discrepancies remain in peak amplitudes and in regions containing weaker or closely spaced hotspots.

\subsection{Training stability and network robustness}

\begin{figure}[t]
    \centering
    \includegraphics[width=\textwidth]{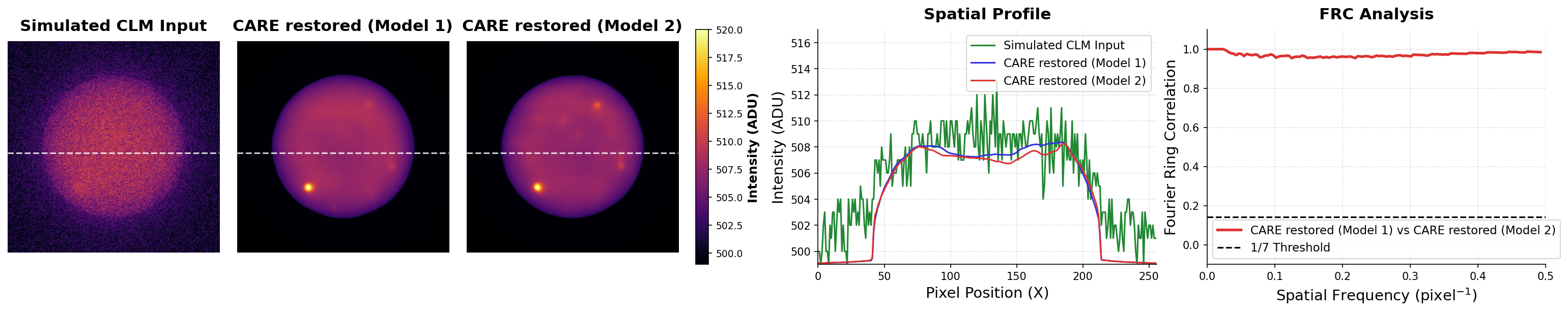}
    \caption{Training stability across independent weight initializations.
    Left columns: a representative simulated CLM image from the validation dataset and the corresponding outputs of two CARE models trained independently from different random weight initializations (Model 1, Model 2). Center: 1D intensity profiles along the dashed line for the simulated CLM input (green) and the two predictions (blue, red).
    Right: FRC curve computed between the two model predictions.}
    \label{fig:training_stability}
\end{figure}

\begin{figure}[t!]
    \centering
    \begin{minipage}[c]{0.64\textwidth}
        \textbf{(a)} \\
        \includegraphics[width=\textwidth]{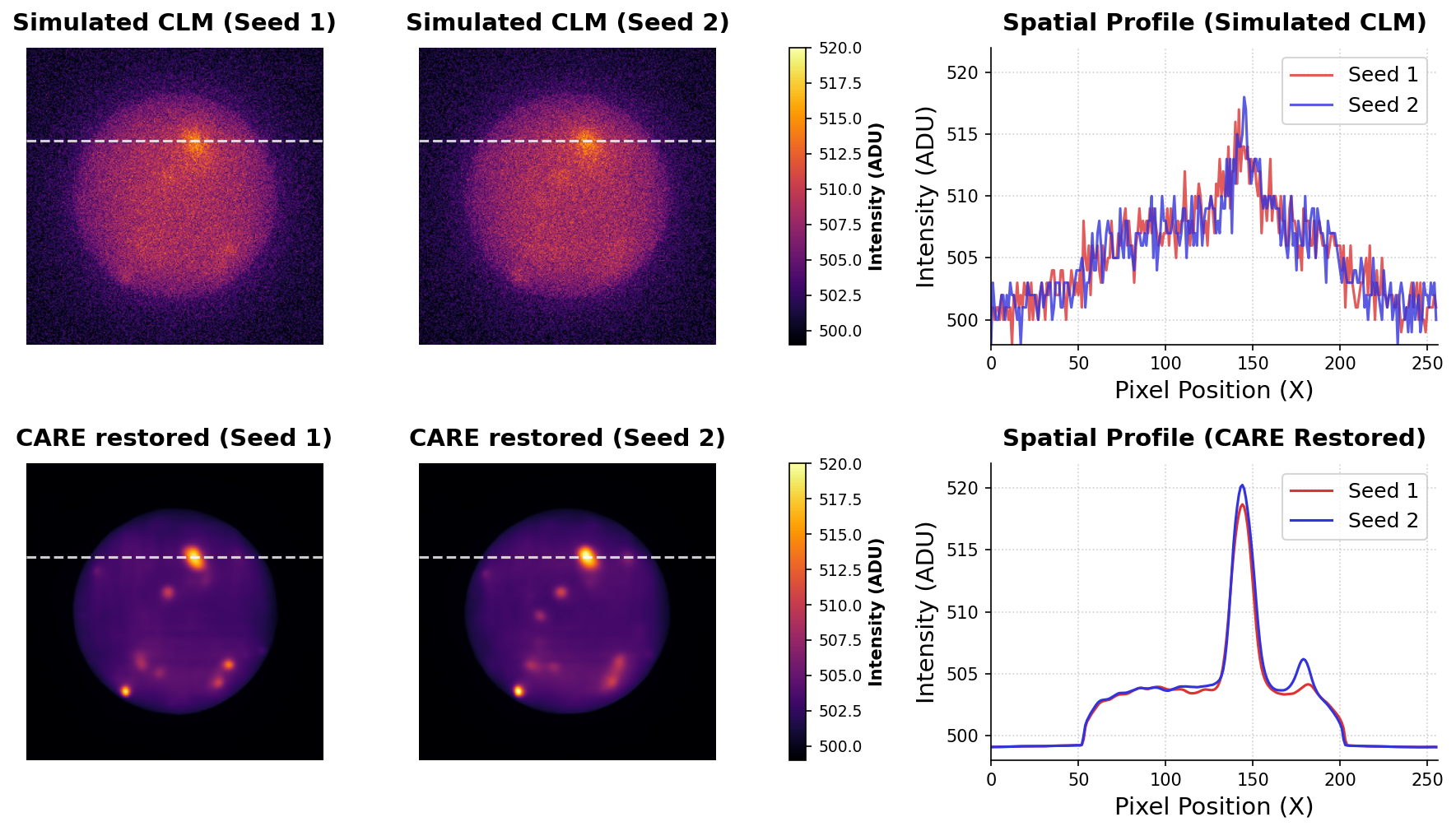}
    \end{minipage}
    \hfill
    \begin{minipage}[c]{0.34\textwidth}
        \textbf{(b)} \\
        \includegraphics[width=\textwidth]{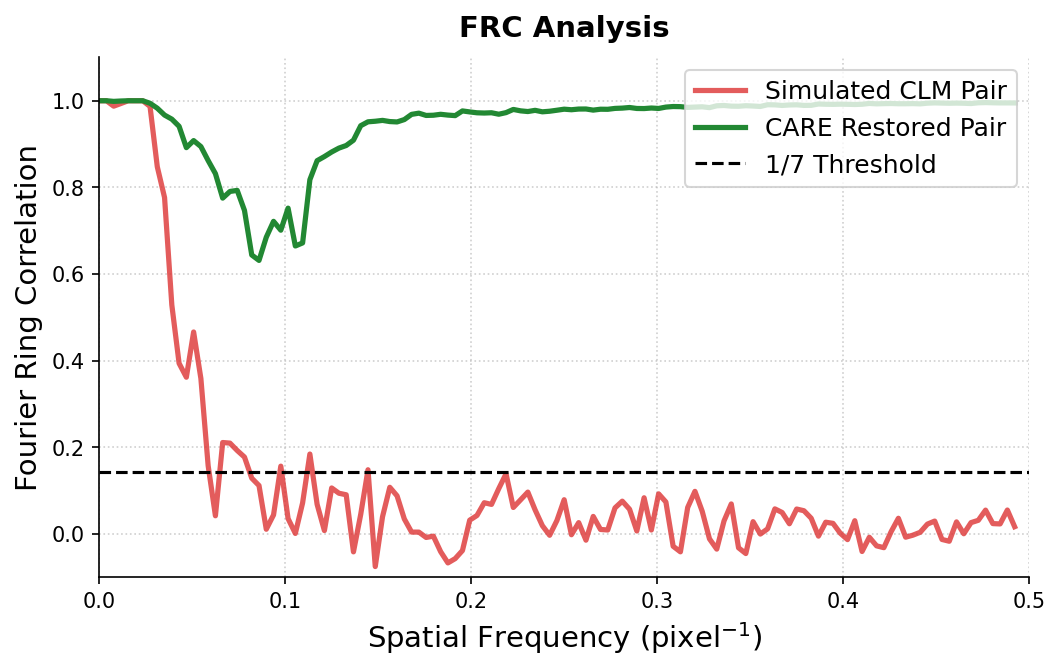}
    \end{minipage}
    \caption{Network robustness across independent noise realizations.
    (a) A representative spheroid simulated twice with identical physical parameters but independent random noise seeds. Top row: the two simulated CLM images (Seed 1, Seed 2) and their 1D intensity profiles along the dashed line. Bottom row: the corresponding CARE restored outputs and their profiles. 
    (b) FRC curves computed between the simulated CLM pair (red) and the CARE restored pair (green).}
    \label{fig:noise_robustness}
\end{figure}

To evaluate training stability, a second CARE model was trained independently using a different random weight initialization, while maintaining the same hyperparameters, training dataset and validation split (training and results of the second model are reported in the Supplementary Material). The predictions of both models were compared on the synthetic validation dataset using SSIM and FRC. Fig.~\ref{fig:training_stability} presents a representative example, showing closely overlapping intensity profiles and a FRC curve remaining above the $1/7$ threshold across the analyzed spatial frequency range. Across the full validation dataset, the mean SSIM between the two model predictions was $0.987 \pm 0.007$, while the corresponding FRC curves remained above the $1/7$ threshold throughout the analyzed frequency range.\\
Noise robustness was evaluated using pairs of simulated CLM images generated from identical spheroid geometries with different random seeds, producing distinct photon statistics. The same trained CARE model was applied independently to both simulated CLM images, and SSIM and FRC were computed before and after restoration.\\
Fig.~\ref{fig:noise_robustness} shows a representative example. The two simulated CLM images exhibit different local intensity fluctuations, as reflected in their 1D intensity profiles. In contrast, the CARE predictions show consistent recovery of hotspot positions and overall peak patterns, as observed in both the images and the corresponding profiles. However, differences in the reconstructed amplitudes and shapes of individual hotspots remain.\\
Across the validation dataset, the mean SSIM between the raw noisy pairs was $0.137 \pm 0.074$, increasing to $0.977 \pm 0.014$ between the corresponding CARE predictions. FRC analysis on the validation dataset showed that the raw image pairs crossed the $1/7$ threshold at a mean spatial scale of $14.2 \pm 3.3$ pixels ($45.4 \pm 10.4~\mu$m), whereas the FRC curves of all restored pairs remained above the threshold throughout the analyzed frequency range.

\subsection{CARE vs N2V on synthetic data}

\begin{figure}[!t]
\includegraphics[width=\textwidth]{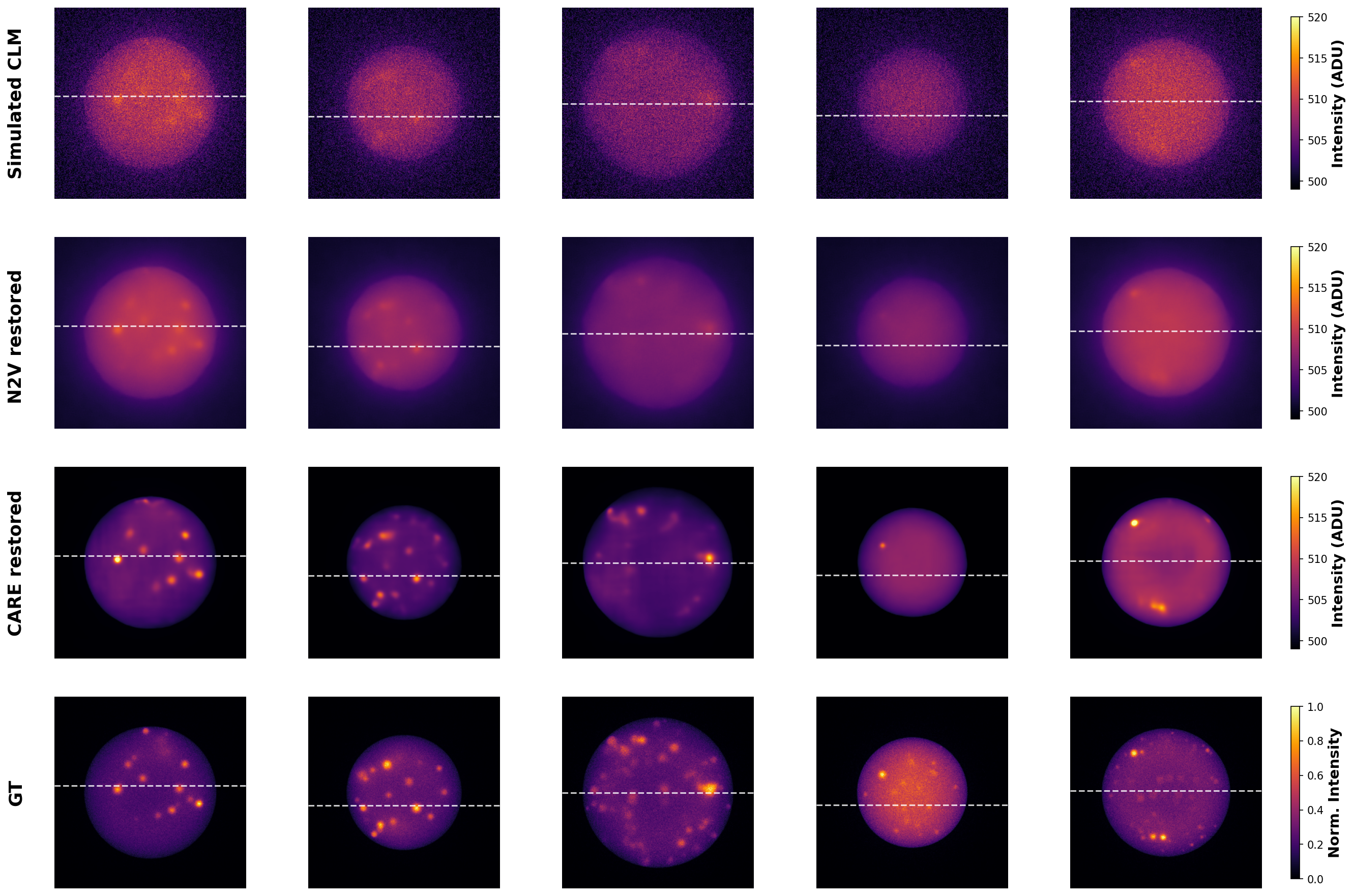}
\vspace{2mm}
\includegraphics[width=0.95\textwidth]{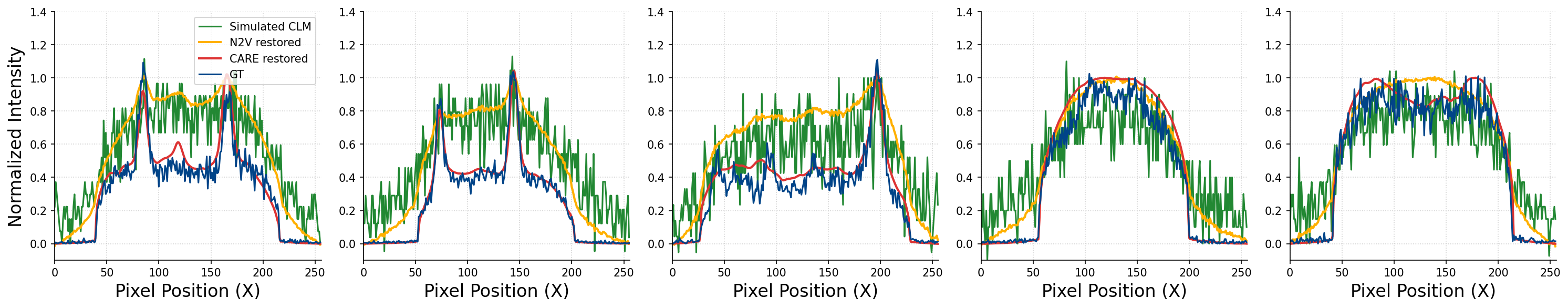}
\caption{Comparison of N2V and CARE restoration using the synthetic dataset. Top panel: representative outputs for five spheroid geometries. Rows display (top to bottom) the simulated CLM, the N2V prediction, the CARE prediction and the GT. Dashed white lines indicate the cross-sections used for 1D profiling.
Bottom panel: normalized 1D intensity profiles along the corresponding cross-sections, comparing the simulated CLM (green), N2V prediction (yellow), CARE prediction (red) and GT (blue). The intensity profiles were normalized using their respective 1st and 99th percentiles.}
\label{fig:care_vs_n2v_synthetic}
\end{figure}

\begin{table}[t]
\centering
\caption{Global restoration metrics on the synthetic dataset. MSE and MAE were computed after independent min-max $[0, 1]$ image normalization.}
\label{tab:care_vs_n2v_metrics}
\vspace{0.2cm}
\begin{tabular}{lcc}
\toprule
\textbf{Comparison} & \textbf{MSE $\pm$ Std Dev} & \textbf{MAE $\pm$ Std Dev} \\
\midrule
Simulated CLM vs GT & $0.0725 \pm 0.0239$ & $0.2488 \pm 0.0486$ \\
N2V restored vs GT       & $0.0492 \pm 0.0239$ & $0.1720 \pm 0.0473$ \\
CARE restored vs GT      & $0.0051 \pm 0.0088$ & $0.0307 \pm 0.0334$ \\
\bottomrule
\end{tabular}
\end{table}

The simulated CLM dataset was also used to train the self-supervised N2V network. Training stopped at epoch 143, with the checkpoint from epoch 123 selected based on the lowest validation loss, which was 0.1117 (training and results of the N2V model applied to the synthetic dataset are reported in the Supplementary Material).\\
Fig.~\ref{fig:care_vs_n2v_synthetic} compares the simulated CLM images, N2V restorations, CARE restorations and corresponding GT images for five representative geometries from the synthetic validation dataset. The N2V outputs show reduced intensity fluctuations compared with the noisy inputs, while largely preserving their underlying broad spatial intensity distributions. In contrast, the CARE reconstructions distinguish localized hotspots more clearly, with spatial profiles that reproduce several of the sharper peaks and intensity gradients observed in the corresponding GT images. The 1D profiles further highlight this difference: N2V reduces noise but retains broad intensity patterns, whereas CARE better resolves individual hotspots and their spatial separation. Nevertheless, differences in peak amplitudes and in the recovery of weaker or closely spaced structures remain evident.\\
These observations are consistent with the quantitative metrics computed across the synthetic validation dataset (Table~\ref{tab:care_vs_n2v_metrics}). After independent min-max normalization of each image to the $[0,1]$ range, the simulated CLM inputs yielded an MSE of $0.0725 \pm 0.0239$ and an MAE of $0.2488 \pm 0.0486$. For N2V, these values decreased to $0.0492 \pm 0.0239$ and $0.1720 \pm 0.0473$, respectively. CARE achieved the lowest errors, with an MSE of $0.0051 \pm 0.0088$ and an MAE of $0.0307 \pm 0.0334$.\\

\subsection{Application to experimental CLM}

\begin{figure}[!t]
\includegraphics[width=\textwidth]{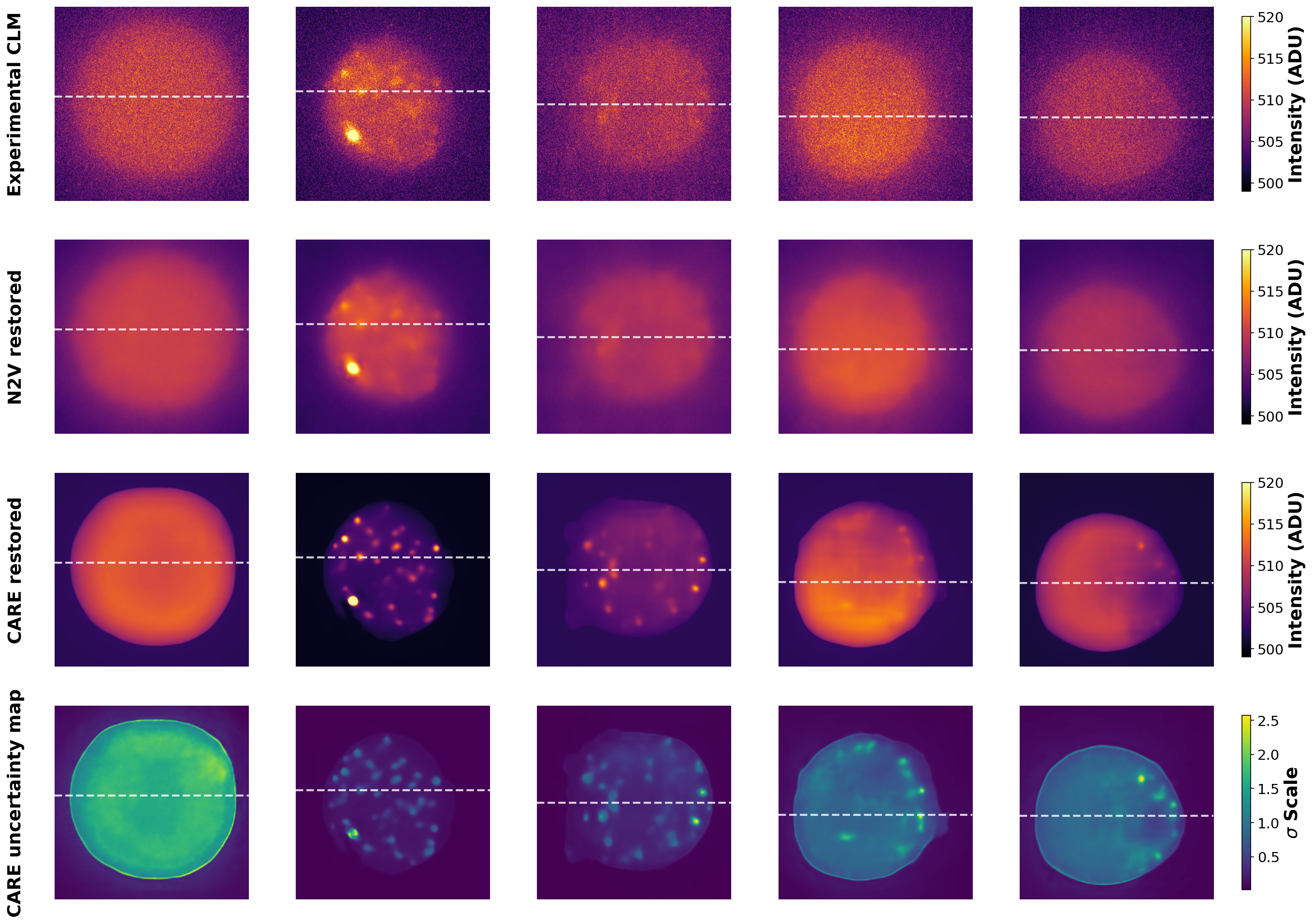}
\vspace{2mm}
\includegraphics[width=0.935\textwidth]{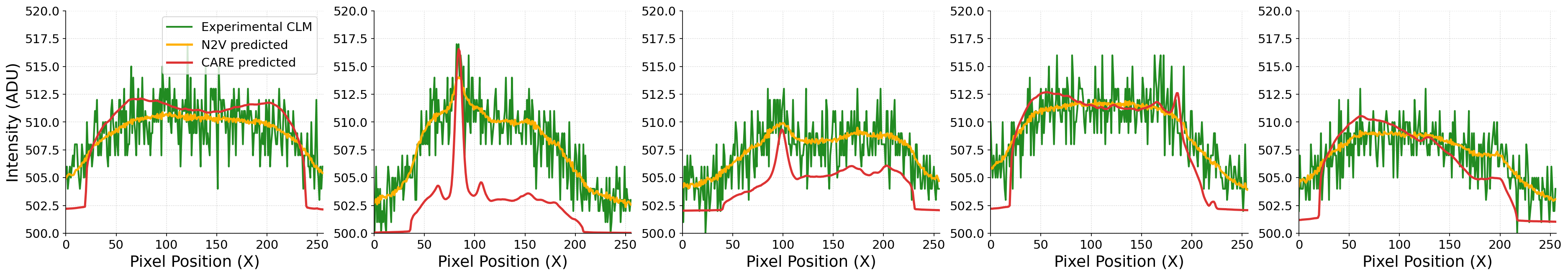}
\caption{Comparison of N2V and CARE restoration on experimental CLM images of U87 tumor spheroids. Top panel: representative outputs for five experimental CLM acquisitions. Rows display, from top to bottom, the experimental CLM images, the N2V predictions, the CARE predictions and the corresponding CARE uncertainty maps ($\sigma$). Dashed white lines indicate the cross-sections used for 1D profiling. Bottom panel: 1D intensity profiles along the corresponding cross-sections, comparing the experimental CLM (green), N2V prediction (yellow) and CARE prediction (red).}
\label{fig:experimental_results}
\end{figure}

The experimental CLM dataset was used to train the self-supervised N2V network. Training stopped at epoch 200, with the checkpoint from epoch 142 selected based on the lowest validation loss, which was 0.2958 (training and results of the N2V model applied to the experimental dataset are reported in the Supplementary Material).\\
The restoration methods were evaluated on the 46 experimental CLM frames acquired from the U87 spheroids (Fig.~\ref{fig:experimental_results}). The five representative examples and their corresponding 1D intensity profiles illustrate differences in the spatial distribution of the signal after processing. N2V reduced high-frequency intensity fluctuations, producing smoother images while generally preserving broad intensity distributions. In contrast, CARE produced more localized intensity patterns in several examples, with some hotspots more clearly distinguishable from their surroundings than in the experimental CLM and N2V images. The 1D profiles further illustrate these differences: while N2V generally retained broad intensity patterns, CARE produced more distinct peaks or sharper intensity transitions. The CARE uncertainty maps showed spatially heterogeneous predicted uncertainty, particular near hotspots location.\\
To quantify local contrast and signal-to-noise characteristics, candidate hotspots were detected exclusively from the CARE predictions. At least one candidate hotspot was identified in 27 of the 46 experimental frames, whereas the remaining 19 frames did not meet the predefined detection criterion. The subsequent quantitative analysis was therefore restricted to these 27 frames. For each frame, the LCR and local SNR values were averaged across the detected hotspots to obtain one value per metric per frame. The same CARE-derived hotspot coordinates and corresponding ROIs were used to calculate both metrics in the experimental CLM, N2V predicted and CARE predicted images, enabling comparison at identical locations.\\
The quantitative results, summarized in Table~\ref{tab:experimental_metrics}, show that the mean LCR was $0.00151 \pm 0.00087$ for experimental CLM images, $0.00056 \pm 0.00064$ for N2V predictions and $0.00469 \pm 0.00302$ for CARE predictions (with corresponding median values of 0.00133, 0.00046 and 0.00415). Similarly, the mean local SNR was $0.370 \pm 0.169$ for experimental CLM images, $0.665 \pm 0.366$ for N2V predictions and $3.239 \pm 1.549$ for CARE predictions (with median values of 0.331, 0.611 and 2.860). Overall, at the hotspot locations identified by the CARE-based detection procedure, CARE yielded substantially higher LCR and local SNR values than both the raw experimental acquisitions and N2V predictions, whereas N2V increased local SNR relative to experimental CLM while reducing LCR.

\begin{table}[t]
\centering
\caption{Local metrics on the experimetal dataset. LCR and SNR were computed across the 27 experimental CLM images in which at least one candidate hotspot was detected in the CARE prediction. The same hotspot coordinates and corresponding regions of interest were used for experimental CLM images, N2V predictions and CARE predictions. Values are reported as mean $\pm$ standard deviation of the per-frame averages, with the median given in parentheses.}
\label{tab:experimental_metrics}
\vspace{0.2cm}
\begin{tabular}{lcc}
\toprule
\textbf{Method} & \textbf{LCR} & \textbf{Local SNR} \\
\midrule
experimental CLM & $0.00151 \pm 0.00087$ (0.00133) & $0.370 \pm 0.169$ (0.331) \\
N2V     & $0.00056 \pm 0.00064$ (0.00046) & $0.665 \pm 0.366$ (0.611) \\
CARE    & $0.00469 \pm 0.00302$ (0.00415) & $3.239 \pm 1.549$ (2.860) \\
\bottomrule
\end{tabular}
\end{table}

\section{Discussion}

The present study demonstrates the potential of combining MC simulations and DL to improve CLM images by mitigating the effect of low photon statistics and positron-induced spatial blurring. The main contribution of our work is the development of a physics-based MC simulation framework that generates paired degraded and GT images of heterogeneous 3D tumor spheroids. By explicitly simulating positron transport, Cerenkov photon generation and optical propagation, this approach makes it possible to train a restoration network against any activity distribution that cannot be directly acquired experimentally.\\
The comparison between simulated CLM images and their corresponding GT references shows the spatial degradation due to Cerenkov photon generation and transport. Unlike the direct optical emission used to generate the simulated GT, Cerenkov photons are emitted along the trajectories of beta particles above the Cerenkov threshold. Combined with optical scattering and noise due to limited photon statistics, this spatial spread limits the detection of small or closely spaced hotspots.\\
The CARE reconstructions demonstrate that a supervised network trained on physically paired MC images can recover degraded spatial information. The improved separation of neighboring hotspots and the closer agreement with GT intensity profiles suggest that the network learns a restoration mapping extending beyond conventional noise suppression. Despite the substantial degradation of the simulated acquisitions, CARE recovered different spatial features that were poorly distinguishable in the noisy images. \\
The stability and noise robustness analyses provide evidence of the reproducibility of the CARE restoration process. The agreement between two independently initialized models suggests that the learned mapping is relatively insensitive to weight initialization under the investigated conditions. Similarly, the high structural similarity between reconstructions obtained from independent noise realizations indicates that CARE can recover consistent spatial patterns despite substantial differences in the noisy images. The FRC results further support this observation, with restored image pairs maintaining spectral correlation above the $1/7$ threshold throughout the accessible frequency range. However, reproducibility should be distinguished from reconstruction accuracy, as independently trained models may share systematic errors originating from the same training data or physical model. Therefore, the agreement with the simulated GT remains the primary evidence of reconstruction accuracy within the synthetic domain.\\
The comparison between CARE and N2V highlights the potential advantage of supervised restoration using MC-generated image pairs over self-supervised denoising. On synthetic data, N2V reduced both the MSE and MAE relative to the noisy inputs. However, the remaining errors were substantially higher than those obtained with CARE and several localized activity peaks remained unresolved after N2V processing. This discrepancy reflects the difference of the two approaches. N2V learns to estimate pixel intensities from their spatial context and is not explicitly trained to reverse positron-induced spatial broadening. CARE, conversely, learns directly from paired degraded and GT images. This enables CARE to address both stochastic intensity fluctuations and the spatial redistribution of the signal, producing a form of physics-guided image restoration that addresses both noise and positron-induced spatial blurring. The substantially lower MSE and MAE obtained with CARE on the synthetic dataset directly support this simulation-driven strategy over more conventional denoising methods.\\
Application to experimental CLM acquisitions revealed different effects of the two restoration approaches on local SNR and LCR. N2V increased the local SNR relative to the raw images while reducing the LCR. This behavior is consistent with denoising that suppresses background fluctuations but can also smooth local intensity gradients, reducing the contrast between peaks and their surroundings. CARE, in contrast, yielded higher values of both metrics at the analyzed hotspot locations. Together with the qualitative observations and intensity profiles, these findings indicate that CARE produces more localized intensity distributions and greater local peak-to-background separation than N2V. The higher local contrast and SNR obtained with CARE suggest that the restoration mapping learned from the MC-generated dataset produces meaningful enhancements in experimental acquisitions, supporting the practical applicability of simulation-driven restoration to CLM.\\
It is useful to consider that the experimental hotspot analysis was performed at locations identified exclusively from the CARE predictions. Applying the same regions of interest to raw, N2V and CARE images ensured a consistent spatial comparison, but the selection procedure may favor structures that are more clearly expressed in the CARE output. Furthermore, the absence of an experimental GT prevents direct verification that the reconstructed peaks correspond to genuine local maxima in the underlying radiotracer distribution. Nevertheless, the analysis provides a quantitative description of how each restoration method modifies local contrast and noise characteristics at the detected hotspot locations.\\
The probabilistic CARE formulation additionally provides spatially varying scale estimates ($\sigma$) alongside the restored image. The heterogeneous distribution of $\sigma$ indicates that the network assigns different levels of predictive uncertainty across the image. One possible contributor is the ambiguity associated with reconstructing a heterogeneous 3D activity distribution from a single 2D optical acquisition. Different source configurations at different depths may produce similar projected signals, making an exact reconstruction inherently ambiguous in some regions, particularly when combined with positron-induced spread and limited photon statistics. \\
A key methodological advantage of the proposed DL approach is the use of MC simulations to generate physically related training pairs while accounting for the heterogeneous geometry and activity distribution of 3D tumor spheroids. Stochastic variations in spheroid dimensions, necrotic core size, hotspot distribution and relative activity, combined with multiple viewing angles and geometric augmentations, produced 15,360 paired images from 80 simulated configurations. Unlike approaches relying on synthetic noise addition or generic degradation, this physics-based supervision establishes a physically motivated relationship between degraded acquisitions and GT images. However, because experimental GT data are unavailable, reconstruction accuracy on biological samples cannot be directly verified. Consequently, CARE predictions should be interpreted as estimates constrained by the simulated training domain rather than definitive representations of the underlying radiotracer distribution. Their reliability therefore depends on the representativeness of the MC forward model. In particular, the optical properties of the spheroids were derived from the literature rather than measured directly on the experimental samples, while the surrounding culture medium and imaging system were described using effective approximations. Differences between simulated and actual imaging conditions could therefore propagate into the learned restoration. Although the stochastic geometries introduce substantial structural variability, the dataset cannot encompass the full biological heterogeneity of real tumor spheroids. Future work should therefore expand the simulated parameter space, directly characterize sample optical properties and validate reconstructed hotspots via complementary imaging or biological measurements to confirm radiotracer uptake variations. Moreover, since experimental evaluation was limited to U87 spheroids under a single CLM setup, generalization to other biological models and imaging systems needs to be investigated.\\
Beyond improvements to the computational model, acquisition modifications could further enhance CLM performance. Alborghetti et al. \cite{alborghetti_cerenkov_2025} previously identified higher numerical aperture objectives and improved sample positioning as potential strategies for increasing photon collection and reducing acquisition-related degradation. Immersion objectives could further improve light collection, while using phenol-red-free culture medium would reduce optical absorption and signal attenuation \cite{alzyoud_vitro_2019}. These hardware optimizations are strictly complementary to computational restoration: while higher photon statistics improve input quality, they cannot eliminate positron-range blurring, making the synergy between optimized acquisition protocols and MC-guided DL essential to maximize recovery accuracy.\\
Beyond CLM, the proposed framework could potentially be adapted to other photon-limited imaging techniques where experimental GT images are difficult to obtain. Adapting the forward model to specific emission and transport mechanisms would allow the generation of paired training datasets for modalities such as RLM and BLM, offering a more general strategy for low-signal imaging.\\
Finally, the proposed approach may support broader applications of CLM in 3D biological models. Unlike BLM, which requires expression of a bioluminescent reporter such as luciferase, CLM exploits optical radiation intrinsically generated by the administered radiotracer and does not require any cell modification to produce a visible signal. This may facilitate its future application, for example, to patient-derived tumor spheroids or organoids, where introducing reporter genes can be impractical or undesirable. Combined with enhanced image quality, CLM could potentially provide a useful readout for investigating spatial heterogeneity in radiotracer uptake across different experimental conditions, including pharmacological or radiotherapy treatments. However, correlating tracer uptake heterogeneity with specific metabolic pathways or treatment efficacy will require dedicated functional assays.\\
In conclusion, this study demonstrates the potential of MC-guided DL to improve CLM of 3D tumor spheroids. Trained exclusively on physics-based synthetic pairs without experimental GT references, CARE achieved substantially lower reconstruction errors than N2V on synthetic data and yielded more localized intensity distributions with superior local contrast and SNR in experimental acquisitions. These results support the feasibility of mitigating both noise and positron-induced spatial blurring through MC-guided restoration. The proposed simulation-driven strategy could also be adapted to other photon-limited imaging modalities, providing a general framework for developing restoration methods when experimental reference images are unavailable.

\FloatBarrier

\bibliography{sample}

\section*{Funding}
The research leading to these results has received funding from AIRC under IG 2022, ID. 27336, project PI: Antonello E. Spinelli. 

\section*{Author contributions statement}
S.P. conceived the study, carried out the data analysis and network training and wrote the manuscript. L.A. performed the experiments, contributed to data analysis and reviewed the manuscript. C.L. contributed to data analysis and reviewed the manuscript. A.E.S. conceived the study, acquired the funding, supervised the experiments, data interpretation and manuscript review.

\section*{Declarations}
\section*{Competing interests}
The authors declare no competing interests.

\end{document}


\begin{center}
    {\textbf{Supplementary Material}}
\end{center}


\begin{figure}[h!]
    
    \makebox[\textwidth][l]{\textbf{(a)}} \\[0mm]
    \includegraphics[width=0.9\textwidth]{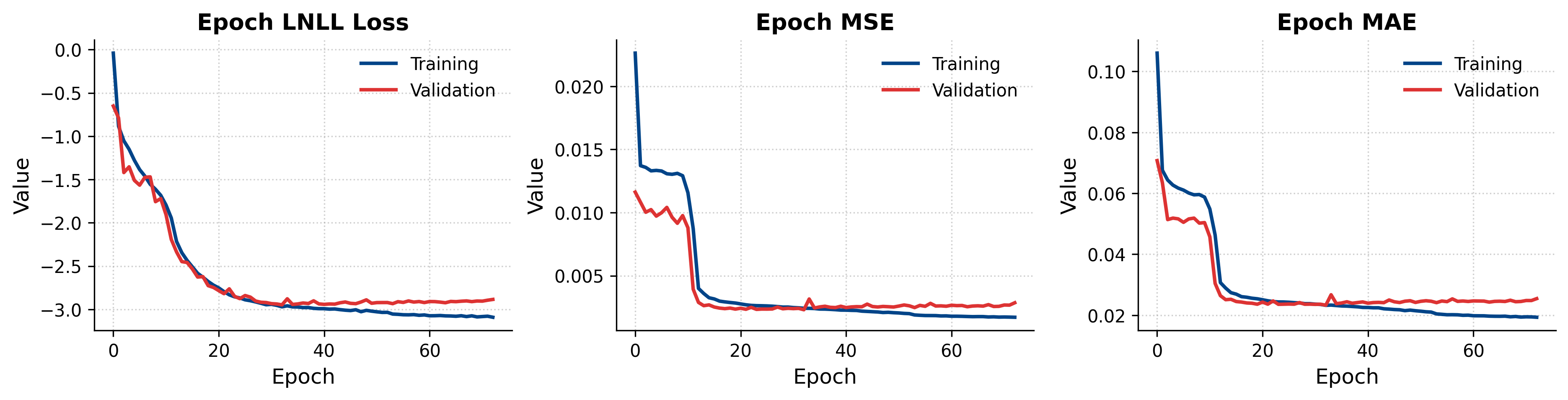} \\[-6mm]
    
    \makebox[\textwidth][l]{\textbf{(b)}} \\[0mm]
    \includegraphics[width=\textwidth]{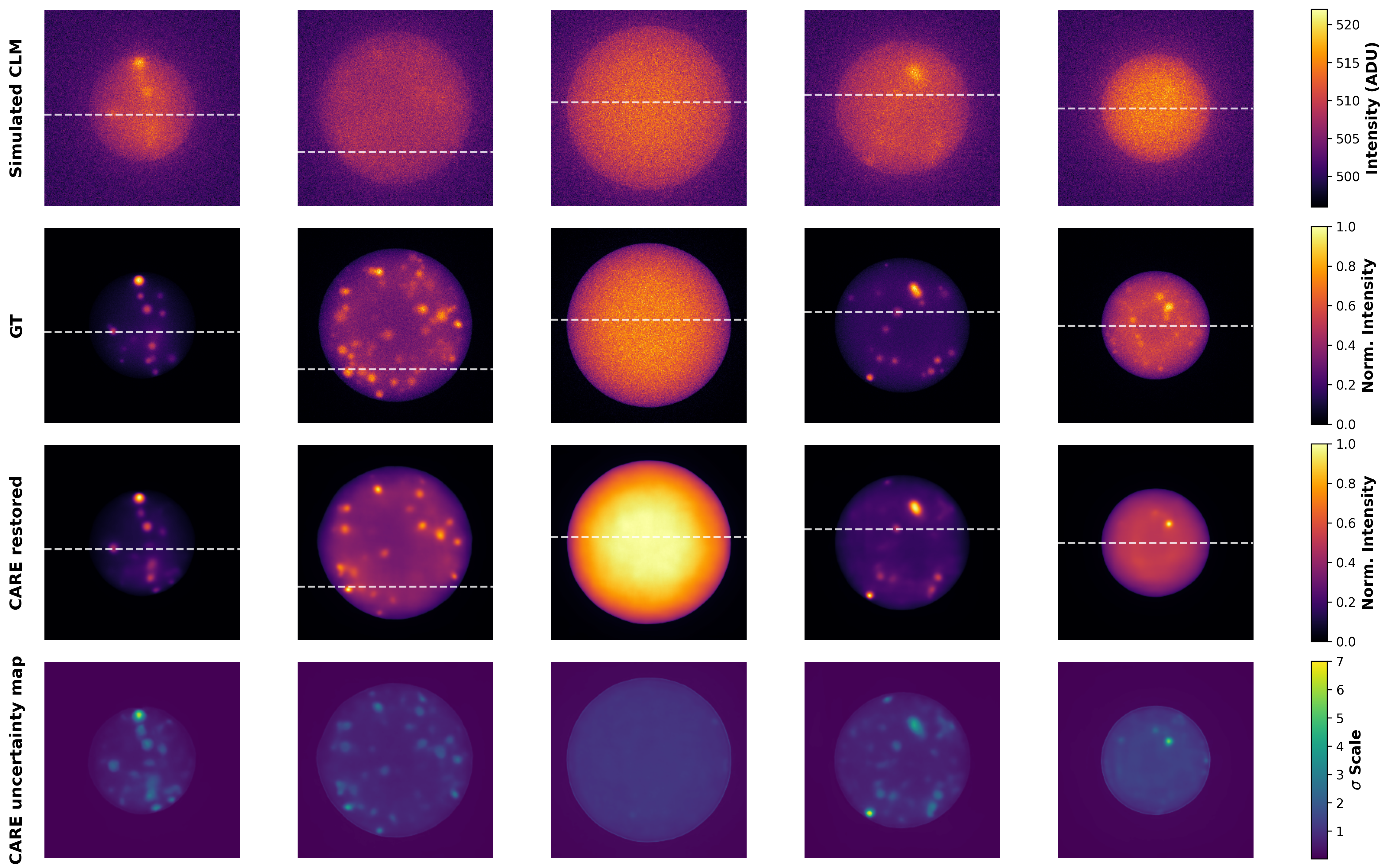} \\[-5mm]
    
    \makebox[\textwidth][l]{\textbf{(c)}} \\[0mm]
    \includegraphics[width=0.93\textwidth]{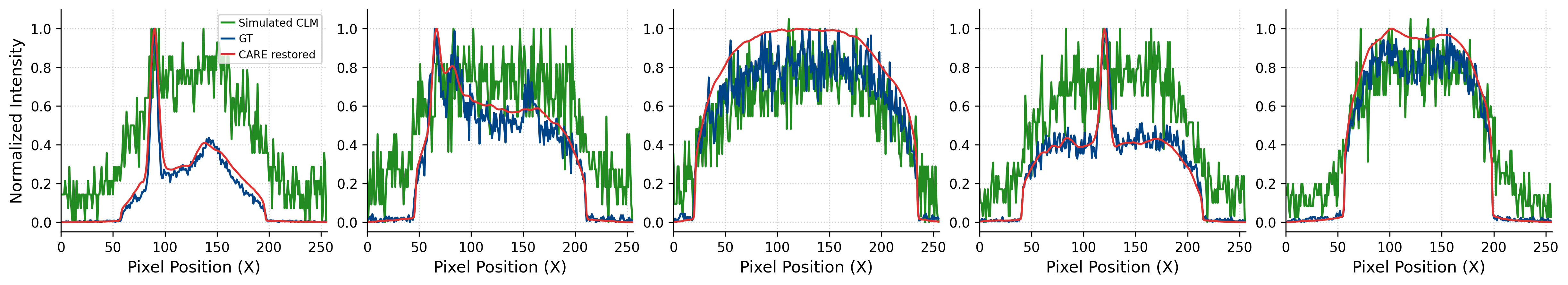}
    
    \caption{Training performance of the second CARE network. It was independently trained using the same synthetic dataset and hyperparameters as the first network, but with a different random weight initialization.
(a) Training and validation metrics (LNLL loss, MSE and MAE) evaluated across epochs.
(b) Restoration results for five representative geometries from the validation dataset. Rows display (top to bottom): simulated CLM images, GT images, CARE predictions and CARE uncertainty maps ($\sigma$). Dashed white lines indicate the cross-sections used for 1D profiling.
(c) 1D intensity profiles comparing the simulated CLM signal (green line), GT signal (blue line) and CARE-restored signal (red line). For visual comparison, all profiles were normalized to the [0, 1] range using their respective 1st and 99th percentiles.}
    \label{fig:supp_care_training_2}
\end{figure}


\begin{figure}[p]
    
    \makebox[\textwidth][l]{\textbf{(a)}} \\[0mm]
    \includegraphics[width=0.9\textwidth]{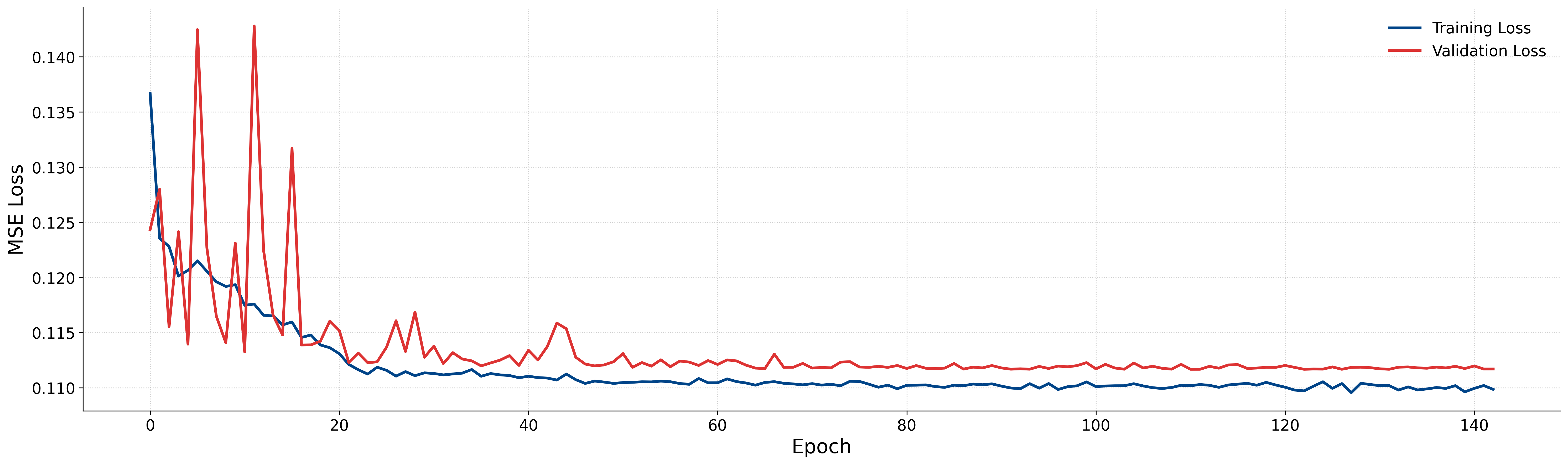} \\[-6mm]
    
    \makebox[\textwidth][l]{\textbf{(b)}} \\[0mm]
    \includegraphics[width=\textwidth]{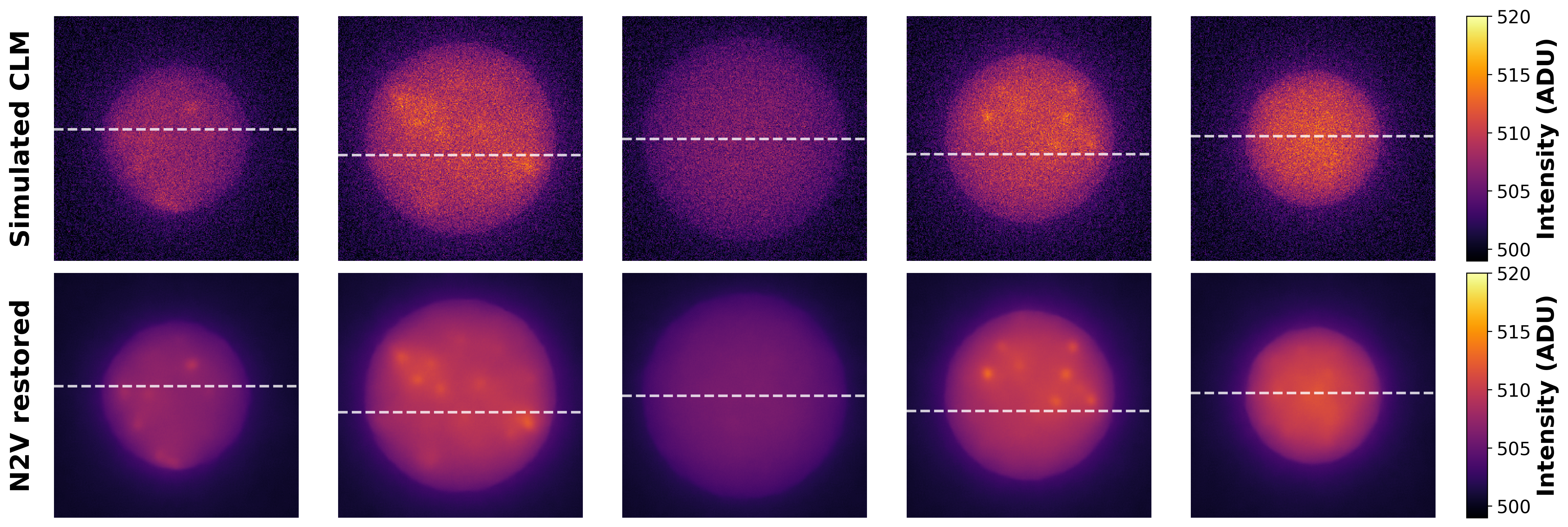} \\[-5mm]
    
    \makebox[\textwidth][l]{\textbf{(c)}} \\[0mm]
    \includegraphics[width=0.93\textwidth]{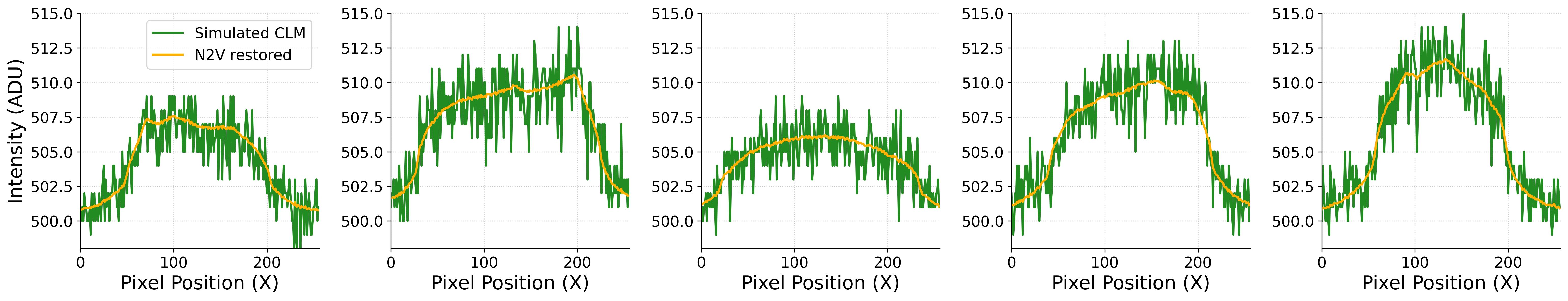}
    
    \caption{Training performance of the N2V network on synthetic CLM images.
(a) Training and validation MSE loss evaluated across epochs.
(b) Results for five representative geometries from the validation dataset. Rows display (top to bottom): simulated CLM images and N2V predictions. Dashed white lines indicate the cross-sections used for 1D profiling.
(c) 1D intensity profiles comparing the simulated CLM signal (green line) and N2V-denoised signal (yellow line).}
    \label{fig:supp_n2v_synthetic}
\end{figure}

\vspace{0.5cm}


\begin{figure}[p]
    
    \makebox[\textwidth][l]{\textbf{(a)}} \\[0mm]
    \includegraphics[width=0.9\textwidth]{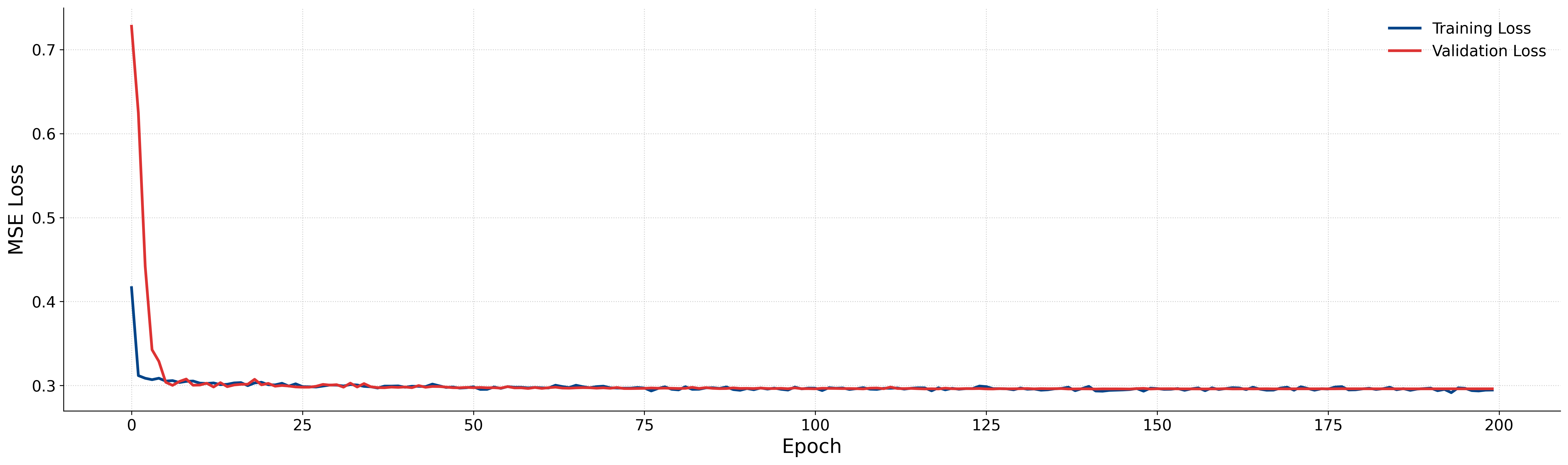} \\[-6mm]
    
    \makebox[\textwidth][l]{\textbf{(b)}} \\[0mm]
    \includegraphics[width=\textwidth]{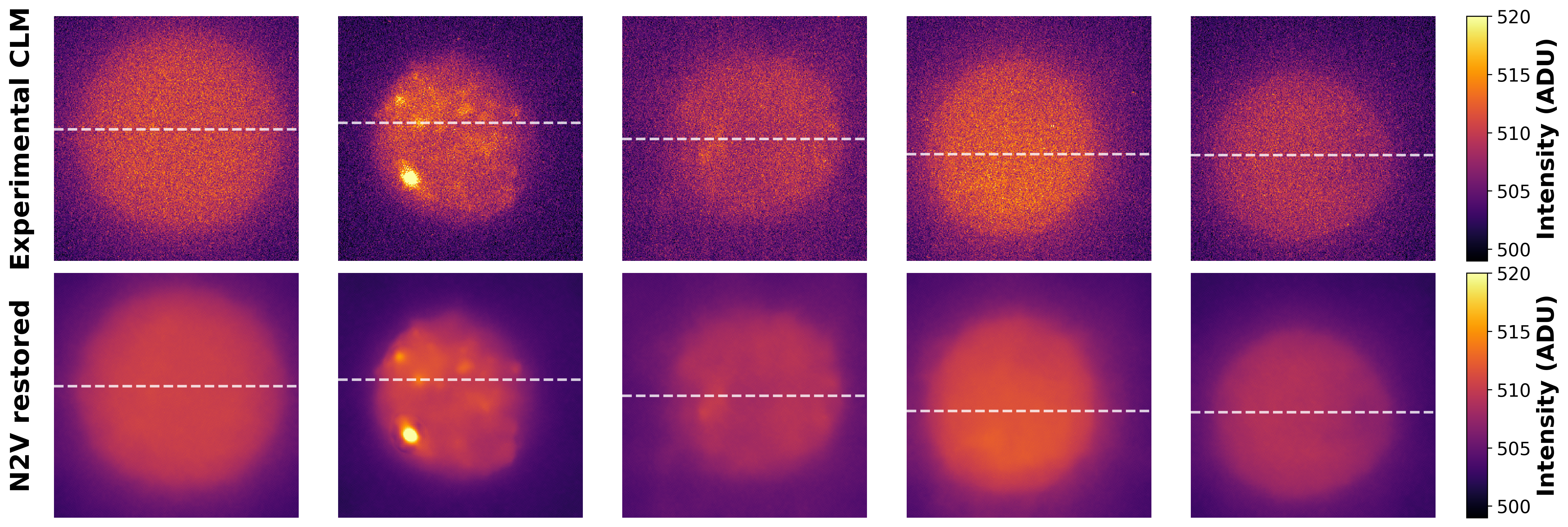} \\[-5mm]
    
    \makebox[\textwidth][l]{\textbf{(c)}} \\[0mm]
    \includegraphics[width=0.93\textwidth]{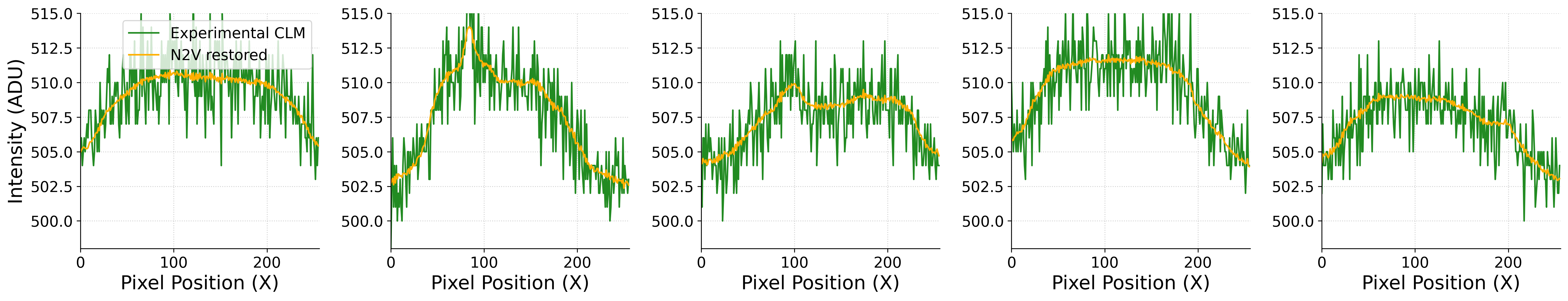}
    
    \caption{Training performance of the N2V network on experimental CLM images.
(a) Training and validation MSE loss evaluated across epochs.
(b) Results for five representative experimental CLM images. Rows display (top to bottom): experimental CLM images and N2V predictions. Dashed white lines indicate the cross-sections used for 1D profiling.
(c) 1D intensity profiles comparing the experimental CLM signal (green line) and N2V-denoised signal (yellow line).}
    \label{fig:supp_n2v_experimental}
\end{figure}